\documentclass[12pt]{iopart}
\usepackage{graphicx}
\begin{document}

\title{ Orbital and spin physics in LiNiO$_2$ and NaNiO$_2$  }

\author{ Albert J W Reitsma,\dag
        \footnote[3]{Present address: Department of Physics, University
         of Strathclyde, John Anderson Building, 107 Rottenrow, Glasgow
         G4 0NG, United Kingdom.}
         Louis Felix Feiner\dag\ddag\
         \\ and Andrzej M Ole\'s\P
}

\address{\dag\  Institute for Theoretical Physics, Utrecht University,
                \\ Leuvenlaan 4, NL-3584 CC Utrecht, The Netherlands
}

\address{\ddag\ Philips Research Laboratories, Prof. Holstlaan 4, \\
                NL-5656 AA Eindhoven, The Netherlands \\
                E-mail: L.F.Feiner@philips.com
}

\address{\P\    Marian Smoluchowski Institute of Physics, Jagellonian
                University, \\ Reymonta 4, PL-30059 Krak\'ow, Poland, \\
                Max-Planck-Institut f\"ur Festk\"orperforschung,
                \\ Heisenbergstrasse 1, D-70569 Stuttgart, Germany \\
                E-mail: A.M.Oles@fkf.mpg.de
}

\begin{abstract}
We derive a spin-orbital Hamiltonian for a triangular lattice of $e_g$
orbital degenerate (Ni$^{3+}$) transition metal ions interacting via
90$^{\circ}$ superexchange involving (O$^{2-}$) anions, taking into
account the on-site Coulomb interactions on both the anions and the
transition metal ions.
The derived interactions in the spin-orbital model are
strongly frustrated, with the strongest orbital interactions selecting
different orbitals for pairs of Ni ions along the three different
lattice directions. In the orbital ordered phase, favoured in mean
field theory, the spin-orbital interaction can play an important role
by breaking the $U(1)$ symmetry generated by the much stronger orbital
interaction and restoring the threefold symmetry of the lattice.
As a result the effective magnetic exchange is non-uniform and
includes both ferromagnetic and antiferromagnetic spin interactions.
Since ferromagnetic interactions still dominate, this offers yet
insufficient explanation for the absence of magnetic order and the
low-temperature behaviour of the magnetic susceptibility of
stoichiometric LiNiO$_2$.
The scenario proposed to explain the observed difference in the
physical properties of LiNiO$_2$ and NaNiO$_2$ includes small
covalency of Ni--O--Li--O--Ni bonds inducing weaker interplane
superexchange in LiNiO$_2$, insufficient to stabilize orbital
long-range order in the presence of stronger intraplane
competition between superexchange and Jahn-Teller coupling.
\end{abstract}




\section{Introduction}
\label{sec:intro}

The low-temperature magnetic behaviour of LiNiO$_2$ has remained
puzzling ever since its peculiar properties were discovered
\cite{Bon57}. For no apparent reason, LiNiO$_2$ does not show magnetic
order nor a cooperative Jahn-Teller effect down to the lowest
temperatures, which is very different from the conventional behaviour
observed in its sister compound NaNiO$_2$.

\begin{figure}
\begin{center}
\vskip 0.5cm
\includegraphics[width=9.7cm,angle=0]{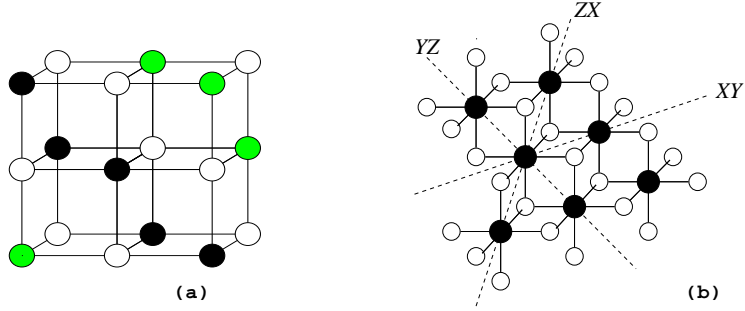}
\end{center}
\vskip -0.4cm
\caption{
Structure of LiNiO$_2$:
(a) Fragment of 3D crystal structure, with  Ni ions shown by filled
circles, Li ions by gray circles, and O ions by open circles.
(b) Nickel layer with adjacent oxygen layers.
Each Ni$^{3+}$ ion (black circles) is surrounded by six oxygens
(open circles). The directions of the Ni$-$Ni bonds are labelled as
$XY$, $YZ$ and $ZX$, corresponding in each case to the plane spanned by
the connecting Ni$-$O bonds.
}
\label{fig:structure}
\end{figure}

Structurally the two systems represent an interesting special category
within the class of correlated transition metal (TM) oxides. LiNiO$_2$
has a layered structure [see figure \ref{fig:structure}(a)]: it is
rhombohedral, consisting of successive (111) planes occupied by Li$^+$,
O$^{2-}$, Ni$^{3+}$, and O$^{2-}$ ions. Thus the Ni$^{3+}$ ions are on a
triangular lattice, with each direct Ni--Ni bond lying along the
diagonal of a nearly square Ni--O--Ni--O plaquette, the Ni--O--Ni bonds
being close to 90 degrees. This is distinct from the more commonly
encountered situation where the bond between two transition metal ions
through the ligand ion connecting them is close to linear (180 degrees),
as e.g. in the perovskites. As pointed out by Mostovoy and Khomskii
\cite{Mos02}, this difference should have important consequences for
the orbital and magnetic superexchange (SE) in LiNiO$_2$, since the SE
within a Ni plane with Ni$^{3+}(t_{2g}^6e_g^1)$ ions
would originate predominantly from virtual charge transfer excitations
$e^12p^6e^1\rightleftharpoons e^22p^5e^1\rightleftharpoons e^22p^4e^2$
along the 90 degrees Ni--O--Ni bonds.

Over the years, a number of experiments (magnetic susceptibility, ESR,
NMR, neutron scattering) have been performed on LiNiO$_2$ samples of
varying stoichiometry \cite{Hir85}--\cite{Cha02}.
From these data one has concluded that
the presence of excess Ni ions in the lithium layers introduces extra
ferromagnetic (FM) coupling between the nickel layers. In addition, for
the samples closest to perfect stoichiometry a positive Curie-Weiss
temperature was found, indicating that the in-plane exchange is also FM.
This is not in accordance with the description given initially by
Hirakawa \etal \cite{Hir85}, namely that LiNiO$_2$ would be a
triangular lattice antiferromagnet (TALAF). Actually, this assumption
was the original motivation for performing magnetic measurements on
LiNiO$_2$, since the TALAF for spin $S=1/2$ is a frustrated system and
the ground state might be some kind of quantum liquid \cite{Faz74}
instead of the classical 120$^{\circ}$ rotated spin arrangement.

At first sight the FM nature of the in-plane correlations is not
surprising when one looks at the three-dimensional (3D) crystal
structure (figure \ref{fig:structure}). As the neighbouring Ni$^{3+}$
ions are connected via two Ni-O-Ni bridges, one has the case of
90$^{\circ}$ SE, which the classical Goodenough-Kanamori-Anderson rules
\cite{Goo63} apparently predict to be FM. In a strong crystal
field the ground state of Ni$^{3+}$ is the low-spin ($t_{2g}^6 e_g^1$)
configuration, so that direct exchange between $t_{2g}$ electrons does
not occur. The exchange interactions between different nickel layers
should also be FM \cite{Rei83}, so that one naively expects a
long-range ordered FM ground state. However, for LiNiO$_2$ no long-range
magnetic order was found down to temperatures very close to $0$ K and
the magnetic susceptibility gradually diverges, giving the impression
that the FM correlations mysteriously disappear. A suggestion was made
by Feiner, Ole\'{s} and Zaanen \cite{Fei97} that the $e_g$ orbital
degeneracy of the Ni$^{3+}$ ion and partly antiferromagnetic (AF)
interactions might be responsible for this peculiar behaviour.

The issue was then addressed by Mostovoy and Khomskii (MK) in an
important paper \cite{Mos02} in which they proposed a realistic
spin-and-orbital model for the Ni plane, which includes the Coulomb
repulsion and the Hund's rule exchange splitting on oxygen. They arrived
at the conclusion that there is a huge degeneracy in the orbital sector,
which is not resolved at the mean-field (MF) level. Yet orbital order
is favoured over an orbital liquid state by the order-out-of-disorder
mechanism, while they claimed that anyway the magnetic interaction is
always FM \cite{Mos02}. From the
absence of an orbital ordered state in LiNiO$_2$ they concluded that the
difference between LiNiO$_2$ and NaNiO$_2$ is probably extrinsic, due to
disorder or electron-lattice interaction. Apparently this has now become
the predominant view, and is as yet not inconsistent with experiments.

However, a conclusion concerning the nature of the magnetic interactions
and the origin of the peculiar properties of LiNiO$_2$ had better be
drawn only after the theoretical prediction for the intrinsic in-plane
behaviour is fully established.
We believe that this is not the case and therefore reanalyze the
situation in this paper. Our finding is that upon inclusion of the
Hund's rule splitting also on the Ni ions and correction of what is
apparently a mistake in the MK analysis, both FM and AF interactions
can occur in the Ni plane, depending upon the orbital arrangement.
Admittedly, this still leaves the difference between LiNiO$_2$ and
NaNiO$_2$ to be explained, but it reopens the case for an intrinsic
mechanism since different orbital phases in the Ni plane could
give rise to different magnetic interactions.

The paper is organized as follows.
In section \ref{sec:formalism} we present the notation used in the
following sections for the spin and orbital operators used to derive
the microscopic model. In section \ref{sec:SEeg} we review some
issues concerning SE, in particular with regard to the Ni--O--Ni
90$^\circ$ bond, and compare this case with the standard situation
encountered for 180$^\circ$ bonds in TM perovskites. The spin-orbital
SE model for the triangular Ni planes of LiNiO$_2$ and NaNiO$_2$ is
presented in section \ref{sec:model}. Next we discuss the strong
frustration of the orbital and spin interactions in this model,
and we investigate its consequences and present possible ground states,
obtained using MF theory for a pair of Ni ions (section \ref{sec:pair})
and for the entire plane (section \ref{sec:GS}).
The implications of the model for the physical properties of LiNiO$_2$
and NaNiO$_2$ are discussed in section \ref{sec:compounds}.
Finally, in section \ref{sec:summ} the main conclusions and a summary
are given. Technical details of the derivation of the model are
presented in \ref{sec:appendix}.

%
%
\section{Pseudospin formalism for degenerate $e_g$ orbitals}
\label{sec:formalism}

For the twofold degenerate $e_g$ orbital state at each site local
operators corresponding to pseudospin $T=1/2$ are introduced, i.e.
$T_i^x$, $T_i^y$ and $T_i^z$, acting as half the Pauli matrices
$\sigma^x, \sigma^y$ and $\sigma^z$ on the two-dimensional (2D)
orbital Hilbert space at site $i$ with basis
\begin{equation}
\displaystyle{1\choose 0}_i\,\equiv |iz\rangle \, \equiv
d_{i,3z^2-r^2}^{\dagger}|0\rangle,
        \hskip 2cm
\displaystyle{0\choose 1}_i\,\equiv |i\bar{z}\rangle \,\equiv
d_{i,x^2-y^2}^{\dagger}|0\rangle.
\label{basis}
\end{equation}
A general superposition is given by
\begin{equation}
\vert i \theta_i \rangle = \cos (\theta_i/2) \, \,
         \displaystyle{1\choose 0}_i
+ \sin (\theta_i/2) \, \, \displaystyle{0\choose 1}_i,
\label{orbital}
\end{equation}
which for example at $\theta_i = \frac{\pi}{3}$ corresponds to a
$d_{z^2-x^2}$ orbital.
The expectation values of the pseudospin operators in
the general orbital state (\ref{orbital}) are
\begin{equation}
\langle T_i^z \rangle = \case{1}{2} \cos \theta_i, \hskip 1cm
\langle T_i^x \rangle = \case{1}{2} \sin \theta_i, \hskip 1cm
\langle T_i^y \rangle = 0 .
\label{expval}
\end{equation}
In order to make the formalism more flexible and to include explicitly
the cubic symmetry of the $e_g$ orbitals, it is convenient to define two
more equivalent basis sets by
\begin{eqnarray}
|i\alpha\rangle \equiv d_{i,3\alpha^2-r^2}^{\dagger}|0\rangle, \hskip 1cm
   \qquad & \left( \{\alpha, \beta, \gamma\} \; \;
                   \textrm{a cyclic} \right.
             \nonumber \\
|i\bar{\alpha}\rangle \equiv d_{i,\beta^2-\gamma^2}^{\dagger}|0\rangle,
          & \left. \; \; \textrm{permutation of\/} \; \;
                   \{x,y,z\} \right)
\label{basisalpha}
\end{eqnarray}
and corresponding rotated pseudospin operators
$I_i^{\alpha}$ and $\bar{I}_i^{\alpha}$ behaving like
$T_i^z$ and $T_i^x$ with respect to those basis sets, i.e.
\begin{eqnarray}
I_i^{x} = - \case{1}{2} T_i^z -\case{\sqrt{3}}{2} T_i^x ,  \hskip 2cm
 & \bar{I}_i^{x} = + \case{\sqrt{3}}{2} T_i^z - \case{1}{2} T_i^x ,
                \nonumber \\
I_i^{y} = - \case{1}{2} T_i^z +\case{\sqrt{3}}{2} T_i^x  ,
 & \bar{I}_i^{y} = - \case{\sqrt{3}}{2} T_i^z - \case{1}{2} T_i^x ,
                \nonumber \\
I_i^{z} = T_i^z , & \bar{I}_i^{z} = T_i^x ,
\label{Iop}
\end{eqnarray}
which satisfy the identities
\begin{equation}
I_i^{x} + I_i^{y}+ I_i^{z} = 0, \hskip 2cm
\bar{I}_i^{x} + \bar{I}_i^{y} + \bar{I}_i^{z} = 0.
\label{identity}
\end{equation}
%
%
%
We can now introduce on-site orbital projection operators by
\begin{equation}
{\mathcal P}_{i}^{\alpha} =
   ( \case{1}{2} {\bi I}_{i} + I_{i}^{\alpha} ) ,  \hskip 2cm
{\mathcal P}_{i}^{\bar{\alpha}} =
   ( \case{1}{2} {\bi I}_{i} - I_{i}^{\alpha} ) ,
\label{orbproj}
\end{equation}
where ${\bi I}_i$ is the unit operator in the 2D orbital Hilbert space
at site $i$.

%
%
We further define two sets of (mutually dependent) orbital-pair
operators, $\{ {\mathcal I}_{ij}^{\alpha \beta},
{\mathcal J}_{ij}^{\alpha \beta}  \}$ and
$\{ {\mathcal V}_{ij}^{\alpha \beta},
{\mathcal W}_{ij}^{\alpha \beta} \}$,
all of which refer to a pair of nearest neighbour TM ions
with their bond $\langle ij \rangle$ lying in the $\alpha \beta$ plane
(see figure \ref{fig:structure}(b)),
\begin{eqnarray}
\label{bondprojI}
{\mathcal I}_{ij}^{\alpha \beta} &=&
   (  {\bi I}_{i} + {\mathcal P}_{i}^{\bar{\alpha}} )
   (  {\bi I}_{j} + {\mathcal P}_{j}^{\bar{\beta}} )
+   (  {\bi I}_{i} + {\mathcal P}_{i}^{\bar{\beta}} )
    (  {\bi I}_{j} + {\mathcal P}_{j}^{\bar{\alpha}} )  \nonumber \\
&=&
   ( \case{3}{2} {\bi I}_{i} - I_{i}^{\alpha} )
   ( \case{3}{2} {\bi I}_{j} - I_{j}^{\beta} )
+   ( \case{3}{2} {\bi I}_{i} - I_{i}^{\beta} )
    ( \case{3}{2} {\bi I}_{j} - I_{j}^{\alpha} ) ,   \\
\label{bondprojJ}
{\mathcal J}_{ij}^{\alpha \beta} &=&
   (  {\bi I}_{i} - {\mathcal P}_{i}^{\bar{\alpha}} )
   (  {\bi I}_{j} - {\mathcal P}_{j}^{\bar{\beta}} )
+   (  {\bi I}_{i} - {\mathcal P}_{i}^{\bar{\beta}} )
    (  {\bi I}_{j} - {\mathcal P}_{j}^{\bar{\alpha}} )  \nonumber \\
&=&
   ( \case{1}{2} {\bi I}_{i} + I_{i}^{\alpha} )
   ( \case{1}{2} {\bi I}_{j} + I_{j}^{\beta} )
+   ( \case{1}{2} {\bi I}_{i} + I_{i}^{\beta} )
    ( \case{1}{2} {\bi I}_{j} + I_{j}^{\alpha} ) ,  \\
\label{bondprojV}
{\mathcal V}_{ij}^{\alpha \beta} &=&
  - {\bi I}_{i} \, ( I_{j}^{\alpha} + I_{j}^{\beta})
  - ( I_{i}^{\alpha} + I_{i}^{\beta} ) \,  {\bi I}_{j}
 =  {\bi I}_{i} \, I_{j}^{\gamma}
    + I_{i}^{\gamma} \, {\bi I}_{j} , \\
\label{bondprojW}
{\mathcal W}_{ij}^{\alpha \beta} &=&
 2 \, ( I_{i}^{\alpha} \, I_{j}^{\beta}
      + I_{i}^{\beta} \, I_{j}^{\alpha} )
 = 2 \, ( I_{i}^{\gamma} \, I_{j}^{\gamma}
        - I_{i}^{\alpha} \, I_{j}^{\alpha}
        - I_{i}^{\beta}  \, I_{j}^{\beta} ) .
\end{eqnarray}
Their expectation values in a pair state
$|i \theta_i \rangle |j \theta_j \rangle$
are given by
\begin{eqnarray}
\label{bondexpI}
\fl  \langle {\mathcal I}_{ij}^{\alpha \beta} \rangle &=&
\case{1}{8} [ 35 + 12 \cos (\theta_{+}+\chi_{\gamma}) \, \cos \theta_{-}
  + 4 \cos^2 (\theta_{+}+\chi_{\gamma}) - 2 \cos^2 \theta_{-} ]  ,   \\
\label{bondexpJ}
\fl  \langle {\mathcal J}_{ij}^{\alpha \beta} \rangle &=&
\case{1}{8} [ 3 - 4 \cos (\theta_{+}+\chi_{\gamma}) \, \cos \theta_{-}
  + 4 \cos^2 (\theta_{+}+\chi_{\gamma}) - 2 \cos^2 \theta_{-} ]  ,    \\
\label{bondexpV}
\fl  \langle {\mathcal V}_{ij}^{\alpha \beta} \rangle &=&
\cos (\theta_{+}+\chi_{\gamma}) \, \cos \theta_{-}            ,      \\
\label{bondexpW}
\fl  \langle {\mathcal W}_{ij}^{\alpha \beta} \rangle &=&
\case{1}{4} [ 4 \cos^2 (\theta_{+}+\chi_{\gamma})
            - 2 \cos^2 \theta_{-} - 1 ]
= \case{1}{4} [ 2 \cos (2 \theta_{+}+ 2 \chi_{\gamma})
                - \cos 2 \theta_{-} ] ,
\end{eqnarray}
where $\theta_{\pm}= (\theta_i \pm \theta_j )/2$, and
$\chi_x = \case{2 \pi}{3}$, $\chi_y = - \case{2 \pi}{3}$, $\chi_z = 0$.

Finally we introduce {\it bond projection operators\/}, needed for
specifying the SE interactions between a pair of TM ions. For the spin
part we will use the familiar projection operators for spin triplet
and spin singlet,
\begin{equation}
{\bi Q}^T_{ij} = \case{3}{4} {\bf 1}_{ij}
   + {\bi S}_i \cdot {\bi S}_j ,       \hskip 2cm
{\bi Q}^S_{ij} = \case{1}{4} {\bf 1}_{ij}
   - {\bi S}_i \cdot {\bi S}_j ,
\label{spinproj}
\end{equation}
where ${\bf 1}_{ij}$ is the unit operator in the four-dimensional (4D)
spin Hilbert space on the bond $\langle ij \rangle$.
For the orbital part we will make use of
\begin{eqnarray}
\label{bondprojO}
{\mathcal Q}_{{\rm O},ij}^{\alpha \beta} &=&
   {\mathcal P}_{i}^{\alpha} {\mathcal P}_{j}^{\beta}
 + {\mathcal P}_{i}^{\beta} {\mathcal P}_{j}^{\alpha} ,  \\
\label{bondprojM}
{\mathcal Q}_{{\rm M},ij}^{\alpha \beta} &=&
   {\mathcal P}_{i}^{\alpha} {\mathcal P}_{j}^{\bar{\beta}}
 + {\mathcal P}_{i}^{\bar{\alpha}} {\mathcal P}_{j}^{\beta}
 + {\mathcal P}_{i}^{\beta} {\mathcal P}_{j}^{\bar{\alpha}}
 + {\mathcal P}_{i}^{\bar{\beta}} {\mathcal P}_{j}^{\alpha} ,  \\
\label{bondprojN}
{\mathcal Q}_{{\rm N},ij}^{\alpha \beta} &=&
   {\mathcal P}_{i}^{\bar{\alpha}} {\mathcal P}_{j}^{\bar{\beta}}
 + {\mathcal P}_{i}^{\bar{\beta}} {\mathcal P}_{j}^{\bar{\alpha}} ,
\end{eqnarray}
where the labelling will be explained in section \ref{sec:model}.
They are conveniently expanded in terms of the operators defined
above, according to (with indices omitted for clarity)
\begin{equation}
\begin{array}{rcrcrcrcrcrcrl}
{\mathcal Q}_{\rm O}
            &=&                     & &
                                    & &             {\mathcal J}
            &=& \case{1}{2} {\bi I} &-& \case{1}{2} {\mathcal V}
                                    &+& \case{1}{2} {\mathcal W} &
  \\
{\mathcal Q}_{\rm M}
            &=&           4 {\bi I} &-& \case{1}{2} {\mathcal I}
                                    &-& \case{3}{2} {\mathcal J}
            &=&             {\bi I} & &
                                    &-&             {\mathcal W} &
   \\
{\mathcal Q}_{\rm N}
            &=&         - 2 {\bi I} &+& \case{1}{2} {\mathcal I}
                                    &+& \case{1}{2} {\mathcal J}
            &=& \case{1}{2} {\bi I} &+& \case{1}{2} {\mathcal V}
                                    &+& \case{1}{2} {\mathcal W} &\! \!,
\end{array}
\label{expQ}
\end{equation}
where ${\bi I} \equiv {\bi I}_{ij}$ now denotes the identity in the
4D orbital-pair Hilbert space.

%
%
\section{Superexchange for degenerate $e_g$ orbitals}
\label{sec:SEeg}

If degenerate $e_g$ orbitals are partly filled, a spin-orbital
Hamiltonian can be constructed from a degenerate-band Hubbard model in
much the same way as one derives the AF Heisenberg model from the
single-band Hubbard model at half-filling.
The Hamiltonian of the $e_g$-band Hubbard model consists of two parts:
there is a hopping term ${\mathcal T}$ modelling transfer of electrons
between nearest-neighbour transition metal sites and a Hubbard (or
Coulomb) term ${\mathcal U}$ describing the on-site interactions. In
the situation where the number of sites equals the number of electrons,
the ground state for ${\mathcal T} = 0$ has twofold spin and twofold
orbital degeneracy on each site. When we allow for a nonzero
${\mathcal T}$ as a small perturbation, the 4$^N$-fold degeneracy is
lifted by virtual electron hopping involving excited states, and the
effective Hamiltonian in second order perturbation theory is
\begin{equation}
\label{ham}
{\mathcal H}_{\rm eff} = \sum_{n \in \! \! \! / \, {\rm G}}
{\mathcal P}_{\rm G} \,
\Big[ {\mathcal T} \vert n \rangle \frac{1}{E_{\rm G} - E_n} \langle n
\vert {\mathcal T} \Big] \, {\mathcal P}_{\rm G} ,
\end{equation}
where ${\mathcal P}_{\rm G}$ is a projection operator onto the ground
state manifold of ${\mathcal U}$. The spin-orbital Hamiltonian is
obtained by writing out (\ref{ham}) in terms of spin and orbital
projection operators at each site $i$.

For a cubic lattice, such as in the perovskites KCuF$_3$ and
K$_2$CuF$_4$, one can describe by this approach also the 180$^{\circ}$
SE, thus treating it formally as direct exchange. The
hopping term is taken as
\begin{equation}
{\mathcal T}=-\bar{t}
 \sum_{\alpha}\,\sum_{\langle ij\rangle\parallel\alpha}\,\sum_{\sigma}
\; d_{i \alpha \sigma}^{\dagger}\:d_{j \alpha \sigma}^{},
\label{Tdirect}
\end{equation}
expressing that hopping is only allowed between nearest-neighbour
directional orbitals
$|\alpha \rangle \equiv d_{3\alpha^2-r^2}^{\dagger}|0\rangle$
oriented along the connecting $\alpha$ axis
($\alpha$ being $x$, $y$ or $z$).
In the following we will therefore call the $\alpha$-orbitals `hopping'
orbitals -- since electrons in these orbitals can hop and contribute to
the kinetic energy,
and similarly call `non-hopping' orbitals the $\bar{\alpha}$-orbitals
($|\bar{\alpha}\rangle \equiv d_{\beta^2-\gamma^2}^{\dagger}|0\rangle$),
orthogonal to the $\alpha$-orbital and oriented perpendicular to the
bond.
In a one-dimensional chain this situation leads to a characteristic
competition between itinerant and localized phases \cite{Dag04}.

The on-site interactions on a TM ion can be represented by \cite{Ole83}
\begin{eqnarray}
\label{Hee}
{\mathcal U}_{\rm TM}&=&
   U \sum_{i\lambda}n_{i\lambda  \uparrow}n_{i\lambda\downarrow}
 +\Big(U-\frac{5}{2}J_{\rm H}\Big)
    \sum_{i,\lambda<\mu}n_{i\lambda}n_{i\mu}
 -2J_{\rm H}\sum_{i,\lambda<\mu}
     {\bi s}_{i\lambda}\cdot {\bi s}_{i\mu}
                                                           \nonumber \\
&+& J_{\rm H}\sum_{i,\lambda<\mu}
 \Big( d^{\dagger}_{i\lambda\uparrow}d^{\dagger}_{i\lambda\downarrow}
       d^{       }_{i\mu\downarrow}d^{       }_{i\mu\uparrow}
      +d^{\dagger}_{i\mu\uparrow}d^{\dagger}_{i\mu\downarrow}
       d^{       }_{i\lambda\downarrow}d^{
}_{i\lambda\uparrow}\Big),
\end{eqnarray}
and are characterized by two parameters: the intraorbital Coulomb
energy $U$ and the exchange energy $J_{\rm H}$.
The interorbital terms in equation (\ref{Hee}) describe electron
interactions between pairs of orthogonal orbitals, i.e., in the
present subspace of $e_g$ orbitals one has $\lambda, \mu \in
\{\alpha, \bar{\alpha}\}$.
The excited states, generated in the virtual $d-d$ transitions and
relevant for SE, have two
electrons on the same ion, and ${\mathcal U}$ contributes a Coulomb
repulsion energy $U$. In the $\bar{t} \ll U$ limit one thus derives an
effective low-energy spin-orbital Hamiltonian with coupling constant
$J_{\rm SE} \propto \bar{t}^{\, 2}/U$,
in which spin and orbital degrees of freedom are interrelated.
By taking also the Hund's rule exchange $J_{\rm H}$ into account
one removes the classical degeneracy of magnetically ordered phases
\cite{Kug82}.
The stable phase at low temperatures has long-range orbital order of a
particular type of mixed orbitals, leading to AF interactions along the
$c$-axis and FM interactions in the $(a,b)$-plane. This ordering was
verified experimentally and has been shown to be stable with respect to
quantum fluctuations for large $J_{\rm H}$ \cite{Ole00}.

In a more realistic treatment of SE one takes into account explicitly
that the hopping takes place via the ligand oxygen ion. The hopping term
is then taken as
\begin{equation}
{\mathcal T}=-t
 \sum_{\alpha}\,\sum_{\langle ij\rangle\parallel\alpha}\,\sum_{\sigma}
\; \Big( d_{i \alpha \sigma}^{\dagger}\:p_{j\alpha\sigma}^{} +
p_{j\alpha\sigma}^{\dagger}\: d_{i \alpha \sigma}^{} \Big),
\label{Tligand}
\end{equation}
and describes charge transfer with amplitude $t \equiv t_{\sigma}$
between the TM-orbital $d_{3\alpha^2-r^2}$ and the oxygen $\sigma$-type
$p$-orbital $p_{\alpha}$, where again the orbitals are oriented along
the connecting $\alpha$-axis. The on-site interaction on oxygen is 
given by
\begin{eqnarray}
\label{Heeo}
{\mathcal U}_{\rm O}&=&
   U_{\rm O}\sum_{j\lambda}n_{j\lambda  \uparrow}n_{j\lambda\downarrow}
 +\Big(U_{\rm O}-\frac{5}{2}J_{\rm O}\Big)
 \sum_{j,\lambda<\mu}n_{j\lambda}n_{j\mu}
 -2J_{\rm O}\sum_{j,\lambda<\mu}
            {\bi s}_{j\lambda}\cdot {\bi s}_{j\mu}
                                                           \nonumber \\
&+& J_{\rm O} \sum_{j,\lambda<\mu}
 \Big( p^{\dagger}_{j\lambda\uparrow}p^{\dagger}_{j\lambda\downarrow}
       p^{       }_{j\mu\downarrow}p^{       }_{j\mu\uparrow}
      +p^{\dagger}_{j\mu\uparrow}p^{\dagger}_{j\mu\downarrow}
       p^{       }_{j\lambda\downarrow}p^{}_{j\lambda\uparrow}\Big),
\end{eqnarray}
with intraorbital Coulomb energy $U_{\rm O}$ and exchange energy
$J_{\rm O}$.
Here the operators $n_{j\lambda\sigma}$, $n_{j\lambda}$ and
${\bi s}_{j\lambda}$ refer to an oxygen ion at site $j$.
For two-hole excitations in the three $2p$ orbitals,
as arises when the SE is derived (see below), this local problem and
the excitation spectrum are isomorphic to those for three $t_{2g}$
orbitals filled by two electrons as in the vanadates \cite{Kha01}.
The effective Hamiltonian is now obtained in fourth order
perturbation theory,
\begin{equation}
\fl {\mathcal H}_{\rm eff} = \sum_{k,l,m \in \! \! \! / \, {\rm G}}
{\mathcal P}_{\rm G} \Big[
{\mathcal T}
\vert k \rangle \frac{1}{E_{\rm G} - E_k} \langle k \vert
{\mathcal T}
\vert l \rangle \frac{1}{E_{\rm G} - E_l} \langle l \vert
{\mathcal T}
\vert m \rangle \frac{1}{E_{\rm G} - E_m} \langle m \vert
{\mathcal T} \Big] {\mathcal P}_{\rm G} .
\end{equation}

In the case of a 180$^{\circ}$ TM--O--TM bond $\langle ij\rangle$ this
is unproblematic. For the so-called $U$-term (Anderson or delocalization
process) \cite{Goo63}, where an electron is effectively transferred from
one TM ion to the other TM ion, schematically represented as
$$
e^1p^6e^1 \rightarrow e^1p^5e^2 \rightarrow e^0p^6e^2
          \rightarrow e^1p^5e^2 \rightarrow e^1p^6e^1 ,
$$
one can simply replace $\bar{t}$ by $t_{\sigma}^2/\Delta$, where $\Delta$
is the excitation energy for transferring an electron from O to TM,
and so the coupling constant in the effective Hamiltonian becomes
\begin{equation}
J_{U} \propto \frac{t_{\sigma}^4}{\Delta^2} \, \frac{1}{U} .
\label{And180}
\end{equation}
Note that the orbital filled in the second step is necessarily the same,
also as regards spin, as the one emptied in the first step.
The so-called $\Delta$-term (Goodenough process or correlation effect)
\cite{Goo63} involves instead electron transfer of two electrons from
the connecting oxygen ion, one to each of the TM neighbours,
$$
e^1p^6e^1 \, \rightarrow \, e^1p^5e^2 \, \rightarrow \, e^2p^4e^2
              \rightarrow \, e^1p^5e^2 \, \rightarrow \, e^1p^6e^1 .
$$
Here the oxygen $2p^4$ configuration involved has two holes with
opposite spin on the same $\sigma$-type $p$-orbital, giving always the
same intermediate state at the oxygen ion with energy
$2 \Delta + U_{\rm O}$, and the contributions to the effective
Hamiltonian have coupling constant
\begin{equation}
J_{\Delta} \propto \frac{t_{\sigma}^4}{\Delta^2} \,
\Big( \frac{1}{2 \Delta + U_{\rm O}}  - \frac{1}{2\Delta} \Big)
= - \frac{t_{\sigma}^4}{\Delta^2} \,
\frac{U_{\rm O}}{2 \Delta (2 \Delta + U_{\rm O}) } .
\label{Good180}
\end{equation}
The reason for the subtraction of the term $\propto 1/{2\Delta}$ will be
discussed below.
For a 180$^{\circ}$ TM--O--TM bond the two processes (Anderson and
Goodenough) make qualitatively similar contributions to the effective
Hamiltonian (at least for a single $e_g$ electron on each TM ion
\cite{Mos04}) because both involve $\sigma$-type hopping.
As the terms contributed to ${\mathcal H}_{\rm eff}$ have identical
form, they can be formally generated by the second-order formalism of
direct exchange above, even though this models only the process giving
the $U$-term in
the SE. To obtain the coupling constants one can simply add equations
(\ref{And180}) and (\ref{Good180}). As these are generally of the same
order of magnitude, inclusion of the $\Delta$-term is important
quantitatively, and is essential to describe the trend in the strength
of SE within the $3d$ TM series \cite{Zaa85,Zaa87}.

In the case of a 90$^{\circ}$ TM--O--TM bond, as on a triangular
lattice, the situation is very different. In the $U$-term
(Anderson) process the oxygen orbital through which the electron is
being transferred is now $\sigma$-type for one TM ion but $\pi$-type for
the other one. Therefore this process can only occur if it involves a
(deep-lying) $t_{2g}$ orbital,
$$
(t_2^6 e^1) p^6 (t_2^6 e^1) \, \leftrightarrow \,
(t_2^6 e^1) p^5 (t_2^6 e^2) \, \leftrightarrow \,
(t_2^5 e^1) p^6 (t_2^6 e^2) ,
$$
and thus contributes to the effective Hamiltonian a term with coupling
constant
\begin{equation}
J_{U} \propto \frac{t_{\sigma}^2 t_{\pi}^2}{\Delta^2} \,
\frac{1}{U+\Delta_{\rm CF}} ,
\label{And90}
\end{equation}
since the energy of the middle intermediate state is increased by the
crystal field splitting $\Delta_{\rm CF}$, and the hopping parameter for
the $t_{2g}$ orbital is $t_{\pi}$ instead of $t_{\sigma}$.
In the charge transfer terms ($\Delta$ process)
the oxygen $2p^4$ states now involve two holes
on different $p$-orbitals, each of ${\sigma}$-type but with respect to a
different TM neighbour. This implies that the oxygen Coulomb interaction
involved is now the interorbital interaction $U_{\rm O}'\equiv U_p$
instead of the intraorbital interaction $U_{\rm O}=U_{\rm O}'+2J_p$.
Moreover it leads to a singlet-triplet splitting, with energies
$U_{s}=U_p+J_p$ and $U_{t}=U_p-J_p$, where
$J_p \equiv J_{\rm O}$ is the (Hund's rule) exchange on oxygen, and thus
the contributions to the effective Hamiltonian have coupling constants
\begin{equation}
J_{\Delta}^{\pm} \propto \frac{t_{\sigma}^4}{\Delta^2} \,
\Big( \frac{1}{2 \Delta + U_p \pm J_p} - \frac{1}{2\Delta} \Big)
= - \frac{t_{\sigma}^4}{\Delta^2} \,
\frac{U_p \pm J_p}{2 \Delta (2 \Delta + U_p \pm J_p)} .
\label{Good90}
\end{equation}

From equations (\ref{And90}) and (\ref{Good90}) one observes that
{\it the dominant contribution to the SE comes from the $\Delta$-term
\/} (Goodenough process), since $t_{\sigma}^2/t_{\pi}^2\simeq 4$
\cite{Mat72}, while $U+\Delta_{\rm CF}\gg 2\Delta$ and $U_p\simeq 2\Delta$
\cite{Zaa90,Boc92,Sai95}. This was already pointed out in reference
\cite{Goo63}, as well as the fact that the magnetic interaction should
then be FM, as equation (\ref{Good90}) shows, which is one of the famous
Goodenough-Kanamori-Anderson rules. Nevertheless, one should be careful
not to jump to conclusions here, since the implications of the orbital
SE interaction were not fully considered by Goodenough \cite{Goo63}
(the case explicitly considered was the SE between Ni$^{2+}$
($t_{2g}^6e_g^2$ : $^3\! A_2$) ions \cite{Gri71},
where the two $e_g$ orbitals are both occupied by one electron, and no
orbital effects can arise). It is further clear that the dominant
$\Delta$-term (Goodenough contribution)
cannot be represented well as an effective second order direct exchange.
Yet this was attempted in a recent paper \cite{Ver04}: this has the
merit of deriving the most general form of the effective Hamiltonian
purely based on symmetry considerations, but it cannot capture the
dependence on the most relevant parameters $U_p$ and $J_p$ (actually
only the weaker $U$-term SE, dependent upon the splitting of the
intermediate Ni$^{2+}$ configurations, is being described).

\begin{figure}
\begin{center}
\vskip -1cm
\includegraphics[width=11.7cm,angle=-90]{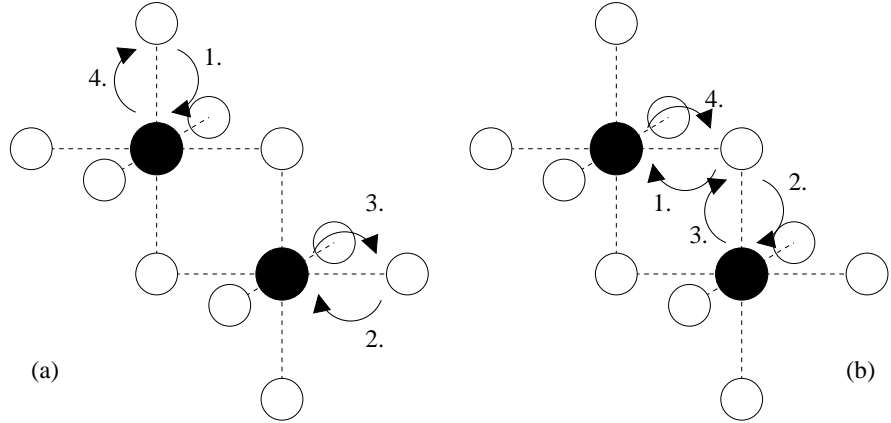}
\end{center}
\vskip -4.5cm
\caption{
Double charge transfer excitations in LiNiO$_2$:
(a) Isolated excitations which renormalize the electronic structure, and
(b) Joined excitations involving a common oxygen site, which contribute
    to the superexchange.
}
\label{fig:renorm}
\end{figure}

A notable peculiarity of the $\Delta$-term, already included in
equations~(\ref{Good180}) and (\ref{Good90}), is that subtraction is
needed of the contribution that would have been obtained if the
electrons transferred to the TM ions would have come from two different
oxygen ions and not from the connecting oxygen ligand, as pointed out by
Mostovoy and Khomskii \cite{Mos04}.
The necessity for this subtraction can be understood as follows.
The reference state (`vacuum') should be considered to be renormalized
by all possible fourth order uncorrelated hopping sequences
[see figure \ref{fig:renorm}(a)], each contributing a term
$t_{\sigma}^4 / ( 2 \Delta^3 )$. In most cases they cancel out because
the three contributions corresponding to hops 2 and 3
in figure \ref{fig:renorm}(a) made along the three cubic axes with
the axis of hops 1 and 4 kept fixed, add up to a constant, since the
three projection operators ${\mathcal P}_i^{\alpha}$ do so because of
equation (\ref{identity}), and thus only add to the vacuum energy.
However, the cancellation fails if the contribution from one axis is
missing because the oxygen ion there is joined with the other TM ion, as
in figure \ref{fig:renorm}(b). The cancellation is restored by adding
this term to the other two and subtracting it from the SE for the pair
of TM ions under consideration.

Note that the above correction implies a sign change of the coupling
constant and so the {\it opposite situation} is favoured than one might
naively expect. In particular, the largest diagonal SE is generally
obtained for the configuration that permits the largest number of
hopping sequences returning to itself, and so this configuration is now
energetically disfavoured instead of favoured. This applies specifically
to purely interorbital SE terms, whereas in spin-spin SE interactions,
which generally originate from the difference of SE for two spin
multiplets, the corrections cancel out and one gets the expected result.

\section{Spin-orbital model for LiNiO$_2$ and NaNiO$_2$}
\label{sec:model}

Let us now reconsider the derivation of the spin-orbital orbital model
for the triangular lattice structure of LiNiO$_2$ \cite{Alb00}, taking
only the charge transfer process ($\Delta$-term) into account, as done
also by MK \cite{Mos02}.
So we consider two nearest neighbour Ni$^{3+}$ ions connected by two
Ni--O--Ni 90$^{\circ}$ bridges, as in figure \ref{fig:structure} or
\ref{fig:renorm}(b), and analyze the fourth order hopping sequences
(which in this order of perturbation theory can be done for each bridge
separately). In order that these virtual excitations lift the ground
state degeneracy it is essential that the intermediate states are
affected by the on-site Coulomb interactions on oxygen and/or nickel,
described by ${\mathcal U}$, see equations (\ref{Hee}) and (\ref{Heeo}).

As indicated above, the relevant states of the oxygen $p^4$
configuration, i.e. those occurring upon hopping, are a triplet
$^3 T_1$ and a singlet $^1 T_2$, denoted for brevity by $t$ and $s$.
They are split by $2 J_p$, while moreover the interorbital Coulomb
repulsion $U_p$ for $p^4$, absent for two $p^5$ configurations, must be
taken into account. The relevant Ni $e_g^2$ states are $^1 \! A_1$,
$^1 \! E$ and $^3 \! A_2$, for which we will use the abbreviations
$S$ (singlet), $D$ (orbital doublet) and $T$ (triplet), respectively.
These terms have equidistant energy levels given by $U+J_{\rm H}$,
$U-J_{\rm H}$ and $U-3J_{\rm H}$, with the triplet being lowest by
Hund's rule, where the Ni$^{2+}$ Coulomb repulsion and Hund's rule
coupling can be expressed in terms of Racah parameters as
$U=A+4B+3C$ and $J_{\rm H}=4B+C$ \cite{Gri71}.

\begin{figure}
\begin{center}
\vskip 0.5cm
\includegraphics[width=11.7cm,angle=-90]{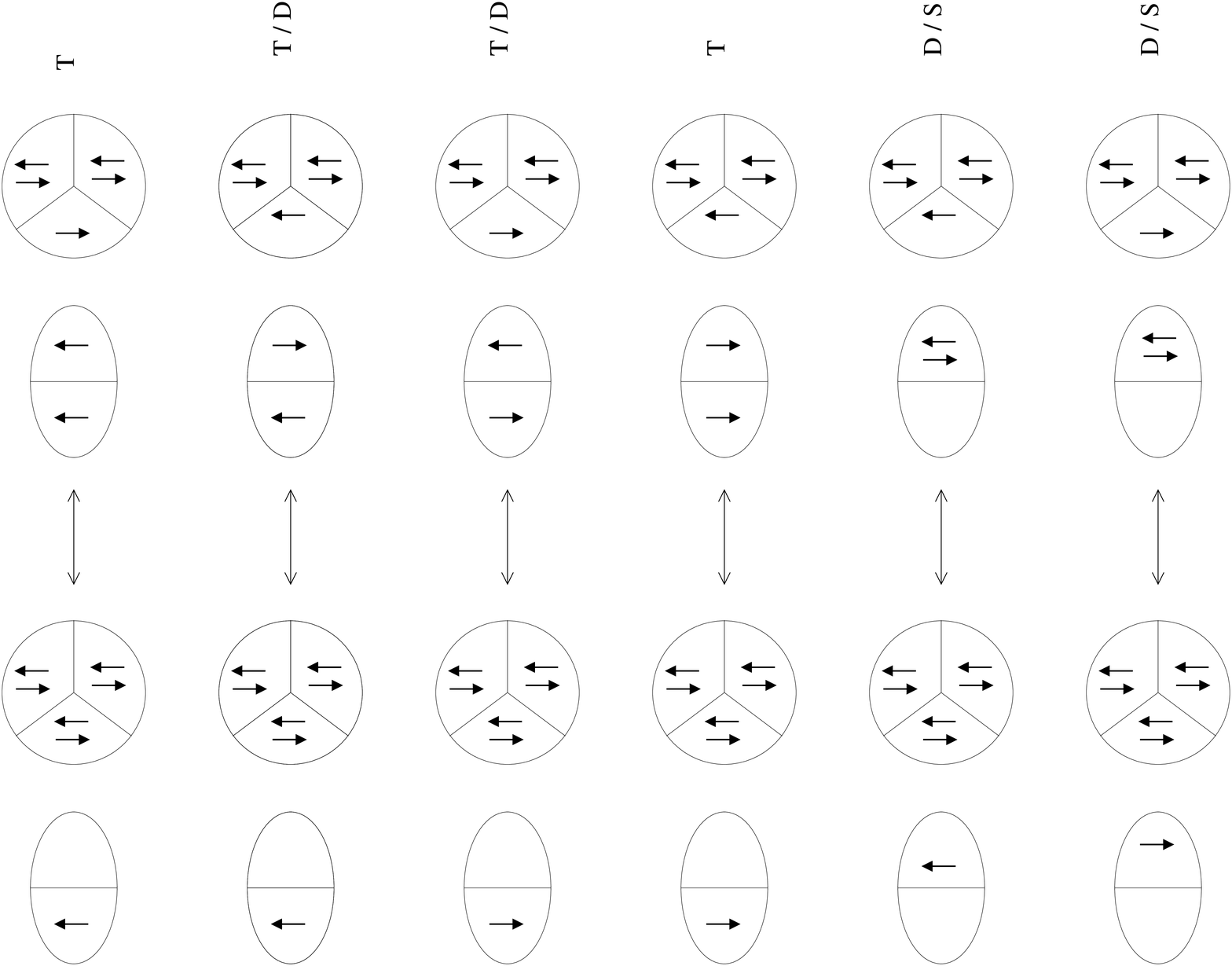}
\end{center}
\caption{
Possible charge excitations from the $e^1 p^6$ to the $e^2 p^5$
configuration. The oval represents the TM ion with one $e_g$ electron,
either in the non-hopping orbital $d_{\beta^2-\gamma^2}$ on the left or
the hopping orbital $d_{3\alpha^2-r^2}$ on the right. The circle
represents the oxygen ion with the electrons in the orbital $p_{\alpha}$
in the segment on the left.
The excited states $^3 \! A_2$, $^1 \! E$ and $^1 \! A_1$ at the TM ion
reached in each case by the $\sigma$-type hopping are
labelled by $T$, $D$ and $S$, respectively.
}
\label{fig:e2p5}
\end{figure}

\begin{figure}
\begin{center}
\vskip 0cm
\includegraphics[width=7.7cm,angle=-90]{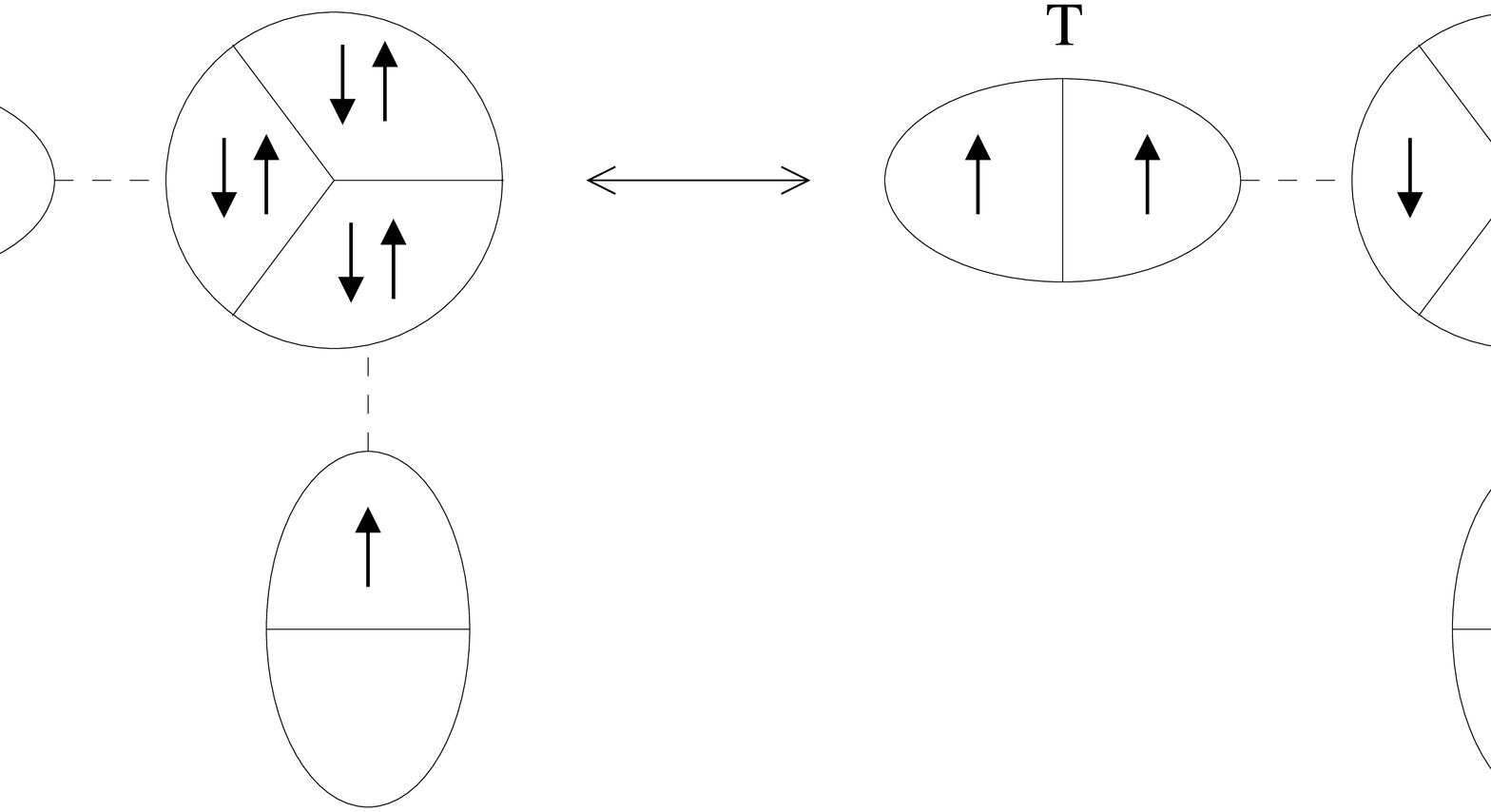}
\end{center}
\vskip -3cm
\caption{
Example of a double charge transfer excitation
$e^1 p^6 e^1 \rightarrow e^2 p^4 e^2$ involved in the superexchange.
The triplet and singlet states at oxygen are labelled by $t$ and $s$;
the labels for the $e_g^2$ states at TM ions are the same as in figure
\ref{fig:e2p5}.
}
\label{fig:doublex}
\end{figure}

The first transition in the hopping sequence is
$e^1 p^6 e^1 \rightarrow e^2 p^5 e^1$, with excitation energy
\begin{equation}
\Delta_{X} = E^{\rm Ni}_{X}(e^2) - E^{\rm Ni}(e^1)
                 + E^{\rm O}(p^5) - E^{\rm O}(p^6),
\label{DeltaX}
\end{equation}
i.e. the charge transfer energy depends upon the Ni$^{2+}$ state
accessed,
\begin{equation}
\Delta_{S}= \Delta + 2 J_{\rm H}, \hskip 1cm
\Delta_{D}= \Delta, \hskip 1cm
\Delta_{T}= \Delta - 2 J_{\rm H},
\label{excNi}
\end{equation}
where it is understood that $U-J_{\rm H}$ has been absorbed into
$\Delta$.
Which states are reached depends on the $e_g$ electron already present
on the Ni$^{3+}$ ion to which the electron hops, as illustrated in
figure~\ref{fig:e2p5}, which shows the six possible hopping channels.
If the $d_{3\alpha^2 - r^2}$ orbital is occupied,
then a hop from the $p_{\alpha}$ oxygen orbital is only possible for
an electron with opposite spin, and the excited states involved are the
spin singlets $^1 \! A_1$ and $^1 \! E$. On the other hand, if the
$d_{\beta^2- \gamma^2}$ orbital is occupied (and therefore the
$d_{3\alpha^2 - r^2}$ orbital is empty), then the spin of the hopping
electron can have either sign, and the $^1 \! E$ and $^3 \! A_2$ states
are reached.
In the second step $e^2 p^5 e^1 \rightarrow e^2 p^4 e^2$, the excitation
energy is raised further by $\Delta_{Y} + U_p \pm J_p$, depending
upon the $e_g^2$ state $Y$ accessed at the other Ni ion and the $p^4$
state ($s$ or $t$) at the oxygen.
Figure \ref{fig:doublex} shows an example of such a double
charge transfer excitation. In the third and fourth step the excitation
is undone, either in the same or in reverse order.

To derive the complete spin-orbital Hamiltonian one must list all
possible initial configurations, and for each of them list all possible
hopping sequences that return to the ground state manifold. The initial
and final states are described by means of projection operators, both
for the orbital occupation and for the spin state, defined in section
\ref{sec:formalism}. The orbital bond projection operators specify
whether the `hopping' orbitals on the two Ni$^{3+}$ ions are both
occupied by an electron (a situation denoted by `O'), whether one
hopping orbital is occupied while the electron on the other ion is in
the non-hopping orbital (`M', for mixed), or whether both
non-hopping orbitals are occupied by the two electrons (`N'). Since the
contributions from the two Ni--O--Ni bridges are independent and may
be simply added,
the operators (\ref{bondprojO}--\ref{bondprojN}) are defined to do so
for each $\alpha \beta$ bond direction. Obviously these operators
depend on the bond direction, and so this gives explicit
orbital anisotropy. For specifying the overall spin state (i.e. formed
by the two spins $1/2$ of the two Ni$^{3+}$ ions), we need the familiar
projection operators (\ref{spinproj}) for spin triplet and spin singlet.

Collecting the contributions from all configurations and all hopping
sequences with the help of diagrams like in figures \ref{fig:e2p5} and
\ref{fig:doublex} one then obtains the Hamiltonian in first instance in
the form
\begin{equation}
\fl {\mathcal H}_{\rm eff} = \sum_{\langle ij\rangle} \Big(
  {\mathcal Q}_{{\rm O},ij}^{\alpha \beta}
    [K_{\rm O}^{\rm T} {\bi Q}^{\rm T}_{ij} 
   + K_{\rm O}^{\rm S} {\bi Q}^{\rm S}_{ij} ]
+ {\mathcal Q}_{{\rm M},ij}^{\alpha \beta}
    [K_{\rm M}^{\rm T} {\bi Q}^{\rm T}_{ij} 
   + K_{\rm M}^{\rm S} {\bi Q}^{\rm S}_{ij} ]
+ {\mathcal Q}_{{\rm N},ij}^{\alpha \beta}
    [K_{\rm N}^{\rm T} {\bi Q}^{\rm T}_{ij} 
   + K_{\rm N}^{\rm S} {\bi Q}^{\rm S}_{ij} ]
    \: \Big) ,
\label{HamK}
\end{equation}
where it is understood that the form of the projection operators depends
on the bond direction. In order to separate the spin dependent part from
the purely orbital part this may be rewritten as
\begin{equation}
\fl {\mathcal H}_{\rm eff} = \sum_{\langle ij\rangle} \Big(
     [ J_{\rm O}^0 {\mathcal Q}_{{\rm O},ij}^{\alpha \beta}
      +J_{\rm M}^0 {\mathcal Q}_{{\rm M},ij}^{\alpha \beta}
      +J_{\rm N}^0 {\mathcal Q}_{{\rm N},ij}^{\alpha \beta} ]
                             {\bf 1}_{ij} \:
   + \: [ J_{\rm O}^S {\mathcal Q}_{{\rm O},ij}^{\alpha \beta}
         +J_{\rm M}^S {\mathcal Q}_{{\rm M},ij}^{\alpha \beta}
         +J_{\rm N}^S {\mathcal Q}_{{\rm N},ij}^{\alpha \beta} ]
           {\bi S}_i \cdot {\bi S}_j \Big) ,
\label{HamJ}
\end{equation}
with orbital and spin-orbital interactions
\begin{equation}
     J_{\rm L}^0 = \case{3}{4} K_{\rm L}^{\rm T} 
                 + \case{1}{4} K_{\rm L}^{\rm S} ,
        \hskip 1cm
     J_{\rm L}^S = K_{\rm L}^{\rm T} - K_{\rm L}^{\rm S}
        \hskip 1cm
        ({\rm L} = {\rm O}, {\rm M}, {\rm N}).
\label{JinK}
\end{equation}

To see how this works in detail and to compare with the analysis of MK,
let us initially ignore the Hund's exchange splitting on Ni (the full
procedure is explained in more detail in \ref{sec:appendix}).
If the initial configuration is N-type, the electrons hopping from the
different $2p$ orbitals at the common oxygen ion into the empty
hopping orbitals on the Ni neighbouring ions, can do so with their
spins oriented in four ways:
both up or both down, thus leaving the oxygen in the triplet $p^4$
state, or with their spins up-down or down-up, both with equal
probability $1/2$ for leaving the oxygen in the triplet $p^4$ or in 
the singlet $p^4$ state, all in all making three triplet and one 
singlet contribution.
As the $e_g^2$ terms on the Ni ions are all equivalent when $J_H=0$,
this is independent of the initial orientation of the spins in the Ni
non-hopping orbitals. Upon including an overall factor of 4, because
both the two excitation transfers and the two deexcitation transfers
can also be made in reverse order, one obtains
\begin{equation}
     K_{\rm N}^{\rm T} = K_{\rm N}^{\rm S} = 12 [XtX] + 4 [XsX] .
\label{KN}
\end{equation}
Here we denote the fourth-order perturbation expressions by giving a
shorthand notation for the middle intermediate state (using here $X$
instead of $S$, $D$ or $T$, because we do not distinguish between those
states yet). Their values are [compare equation (\ref{Good90}) above]
\begin{equation}
  [XtX] = \frac{t^4}{\Delta^2} \,
        \Big(- \frac{1}{2 \Delta + U_p - J_p} + \frac{1}{2\Delta} \Big)
        = \frac{t^4}{\Delta^2} \,
        \frac{U_p - J_p}{2 \Delta (2 \Delta + U_p - J_p)} ,
\label{XtX}
\end{equation}
\begin{equation}
  [XsX] = \frac{t^4}{\Delta^2} \,
        \Big(- \frac{1}{2 \Delta + U_p + J_p} + \frac{1}{2\Delta} \Big)
        =  \frac{t^4}{\Delta^2} \,
        \frac{U_p + J_p}{2 \Delta (2 \Delta + U_p + J_p)} .
\label{XsX}
\end{equation}

If the initial configuration is M-type, the result is equally
independent of the initial Ni--Ni spin state. Although the spin of one
transferred electron must be opposite to the spin of the electron
occupying the hopping orbital on Ni, the other transferred electron can
have its spin still either parallel to that of the first one, leaving a
triplet $p^4$ state on oxygen, or antiparallel, yielding a triplet or a
singlet with probability $1/2$ each. Including again the factor 4, one
obtains
\begin{equation}
     K_{\rm M}^{\rm T} = K_{\rm M}^{\rm S} = 6 [XtX] + 2 [XsX] .
\label{KM}
\end{equation}

Only if the initial configuration is O-type, the result is different,
because each of the transferred electrons must have its spin opposite to
that of the electron already occupying the hopping orbital on Ni. So, if
the spins of the electrons on Ni are parallel, the initial state thus
being a Ni--Ni spin triplet, then the transferred electrons necessarily
also have parallel spins, leaving oxygen in the triplet $p^4$ state. If
the electrons on the Ni ions have antiparallel spins, i.e. being either
in a spin triplet or in a spin singlet state depending upon the phasing
between up-down and down-up, then the spins of the electrons being
transferred are also antiparallel, and have the same phasing, so again
the Ni--Ni spin triplet yields an oxygen $p^4$ triplet, while the Ni--Ni
spin singlet yields an oxygen $p^4$ singlet. Therefore, upon inclusion
of the factor 4,
\begin{equation}
     K_{\rm O}^{\rm T} = 4 [XtX] , \hskip 1cm
     K_{\rm O}^{\rm S} = 4 [XsX] .
\label{KO}
\end{equation}
It follows that
\begin{eqnarray}
\label{J_0}
     J_{\rm N}^0  = 4 J_T , \hskip 1cm
     J_{\rm M}^0  = 2 J_T , \hskip 1cm
     J_{\rm O}^0  =   J_T , \\
\label{J_S}
 J_{\rm N}^S  = 0 , \; \; \; \; \hskip 1cm
 J_{\rm M}^S  = 0 , \; \; \; \; \hskip 1cm
 J_{\rm O}^S  = - J_{TS} , \\
\bs
\label{J_T}
\fl  J_T \; \: = \; \; \; 3 \, [XtX] + [XsX] \; \;  =
\frac{2 t^4}{\Delta^3} \,
\frac{\Delta ( 2 U_p - J_p ) + U_p^2 - J_p^2 }
{(2 \Delta + U_p)^2 - J_p^2}
\simeq
\frac{2 t^4}{\Delta^3} \,
\frac{ U_p }{2 \Delta + U_p} , \\
\fl  J_{TS} = - 4 ( [XtX] - [XsX] ) =
\frac{2 t^4}{\Delta^2} \,
\frac{4 J_p } {(2 \Delta + U_p)^2 - J_p^2} \hskip 1.2cm
\simeq
\frac{2 t^4}{\Delta^2} \,
\frac{4 J_p } {(2 \Delta + U_p)^2},
\label{J_TS}
\end{eqnarray}
where the final expressions on the right in equations (\ref{J_T}) and
(\ref{J_TS}) are the results in the limit $J_p \ll \Delta, U_p$.
So the effective Hamiltonian is
\begin{eqnarray}
 {\mathcal H}_{\rm eff}^{(0)} &=& \sum_{\langle ij\rangle} \Big(
         J_T [  {\mathcal Q}_{{\rm O},ij}^{\alpha \beta}
            + 2 {\mathcal Q}_{{\rm M},ij}^{\alpha \beta}
            + 4 {\mathcal Q}_{{\rm N},ij}^{\alpha \beta} ]
                                           {\bf 1}_{ij} \:
   - \: J_{TS}\: {\mathcal Q}_{{\rm O},ij}^{\alpha \beta}  \:
           {\bi S}_i \cdot {\bi S}_j \Big)   \nonumber \\
&=& \sum_{\langle ij\rangle} \Big(
         J_T  {\mathcal I}_{ij}^{\alpha \beta} {\bf 1}_{ij} \:
   - \: J_{TS}\: {\mathcal J}_{ij}^{\alpha \beta} \:
           {\bi S}_i \cdot {\bi S}_j \Big) ,
\label{HamJH=0}
\end{eqnarray}
with ${\mathcal I}_{ij}^{\alpha \beta} $ and
${\mathcal J}_{ij}^{\alpha \beta}$ given by (\ref{bondprojI}) and
(\ref{bondprojJ}), and understood to depend upon the direction of the
bond $\langle ij \rangle$.

Comparison with the results of MK \cite{Mos02} shows that our analysis
confirms their finding that the purely orbital interaction is
{\it stronger by one order of magnitude\/} than any spin dependent
interaction, $J_T \gg J_{TS}$, in agreement with the conjecture made
by Reynaud \etal and based on their experimental data \cite{Rey01}.
One recognizes that this comes about because, at $J_{\rm H}=0$,
all spin dependence originates from the singlet-triplet splitting
of the oxygen $2p^4$ configuration, and $J_{TS}$ is therefore smaller by
a factor $\sim J_p/U_p\simeq 0.1$ than the orbital interaction $J_T$.
Our analysis also confirms the form of the orbital interaction as being
$\propto {\mathcal I}_{ij}^{\alpha \beta}$.
However, we find a \textit{different} form for the orbital dependence of
the mixed spin-orbital interaction, since MK give
${\mathcal I}_{ij}^{\alpha \beta}$ instead of
${\mathcal J}_{ij}^{\alpha \beta}$.
Apparently their result is incorrect, because the reasoning above
clearly demonstrates that only the O-type configuration can give
spin dependence, and this is what is expressed by the operator
${\mathcal J^{\alpha \beta}}={\mathcal P_{\rm O}^{\alpha \beta}}$.
The difference is important, because the bounds on the expectation
values are very different,
$\case{49}{8} \ge \langle {\mathcal I}^{\alpha \beta} \rangle \ge
\case{25}{8}$ whereas
$\case{9}{8} \ge \langle {\mathcal J}^{\alpha \beta} \rangle \ge 0$
[see equations (\ref{bondexpI}) and (\ref{bondexpJ})],
and so MK concluded that the spin exchange is effectively always
FM for any orbital state, while we conclude that this is at best
marginally so, i.e. the spin exchange interaction on a bond
$\langle ij \rangle$ given by (\ref{HamJH=0}) can vanish for some
specific combination of orbital states at sites $i$ and $j$.

Moreover, the latter conclusion is only valid up to the present level
of approximation, where only the Coulomb interactions on oxygen have
been taken into account.
If we include also the Coulomb interaction on nickel, i.e. allow 
$J_{\rm H}$ to be finite, and consider the leading correction to 
(\ref{HamJH=0}), we find that the spin exchange may even become 
weakly AF.
To see this we first observe, as demonstrated in \ref{sec:appendix},
that quite generally the coupling constants satisfy
\begin{equation}
     |J_{\rm O}^S| \gg J_{\rm M}^S \gg |J_{\rm N}^S| \hskip 0.5cm
       {\rm with} \; J_{\rm M}^S > 0 ,
\label{Jrelation}
\end{equation}
so that the next spin-dependent term to be considered is
\begin{equation}
{\mathcal H}_{\rm eff}^{(1)} =
 J_{\rm M}^{S} \: \sum_{\langle ij\rangle}
    {\mathcal Q}_{{\rm M},ij}^{\alpha \beta}  \:
           {\bi S}_i \cdot {\bi S}_j
= J_{\rm M}^{S} \: \sum_{\langle ij\rangle}
 \Big[ {\bi I}_{ij} - {\mathcal W}_{ij}^{\alpha \beta} \Big] \:
           {\bi S}_i \cdot {\bi S}_j  ,
\label{Ham1}
\end{equation}
and next that it follows from equation (\ref{bondexpW}) that
$\case{7}{4} \ge \langle {\bi I} - {\mathcal W}^{\alpha \beta} \rangle
\ge \case{1}{4}$. We will investigate below under which conditions an AF
net spin exchange on a bond $\langle ij\rangle$ could actually occur.

For that purpose it is useful, when allowing both $J_{\rm M}^S$ and
$J_{\rm N}^S$ to be finite, to rewrite the full spin-orbital SE
Hamiltonian (\ref{HamJ}) explicitly as the sum of a purely orbital part
and a spin dependent part,
\begin{equation}
{\mathcal H}_{\rm eff} = {\mathcal H}_{\rm eff,o}
                         + {\mathcal H}_{\rm eff,s} , \\
\label{Hameff}
\end{equation}
\begin{eqnarray}
\label{HamT}
{\mathcal H}_{\rm eff,o}
&= \sum_{\langle i,j\rangle}
   \Big(   \bar{J}'_T {\mathcal I}_{ij}^{\alpha \beta}  \:
         + \bar{J}_T  {\mathcal J}_{ij}^{\alpha \beta}  \: \Big)
                 \:      {\bf 1}_{ij}
 = J_{\tau} \sum_{\langle i,j\rangle}
   {\mathcal W}_{ij}^{\alpha \beta}  \:  {\bf 1}_{ij} ,  \\
\label{HamTS}
{\mathcal H}_{\rm eff,s}
&= \sum_{\langle i,j \rangle}
   \Big( \bar{J}''_{TS} {\bi I}_{ij} \:
        -\bar{J}'_{TS}  {\mathcal I}_{ij}^{\alpha \beta}  \:
        -\bar{J}_{TS}   {\mathcal J}_{ij}^{\alpha \beta}  \: \Big)
\:
                         {\bi S}_i \cdot {\bi S}_j ,  \nonumber \\
&= \sum_{\langle i,j\rangle}
   \Big(  - J_{\sigma} {\bi I}_{ij} \:
          + J_{\nu} {\mathcal V}_{ij}^{\alpha \beta}  \:
          - J_{\mu} {\mathcal W}_{ij}^{\alpha \beta}  \: \Big) \:
                    {\bi S}_i \cdot {\bi S}_j ,
\end{eqnarray}
with the exchange constants all positive and satisfying (see
\ref{sec:appendix})
\begin{equation}
0 < J_{\sigma} < J_{\nu} < J_{\mu} \ll J_{\tau} .
\label{Jinequality}
\end{equation}
Note that $J_{\sigma}$ is the pure spin SE, while $J_{\nu}$ is the
strength of the SE between a spin-and-orbital operator
$
I_{i}^{\alpha}{\bi S}_{i}
$
on one site and a pure spin operator
$
{\bi I}_{j}{\bi S}_{j}
$
on the other site and $J_{\mu}$ is the SE between spin-and-orbital
operators on both sites. So equation (\ref{Jinequality}) shows that the
pure spin SE is actually always the weakest interaction (for the above
case of $J_{\rm H}=0$, one has
$\bar{J}'_{T}=J_{T}$, $\bar{J}_{T}=0$,
$J_{\tau}=\case{1}{2} J_{T}$;
$\bar{J}''_{TS}=\bar{J}'_{TS}=0$,
 $\bar{J}_{TS}=J_{TS}$,
$J_{\sigma}=J_{\mu}=J_{\nu}=\case{1}{2} J_{TS}$).
In the first expression in equation (\ref{HamT}) we have dropped a
constant, while in obtaining the second expression we have used that
$
\sum_{\langle i,j\rangle} {\mathcal V}_{ij}^{\alpha \beta}=0,
$
as follows from equations (\ref{bondprojV}) and (\ref{identity}).
Note that the latter simplification is not possible in equation
(\ref{HamTS}), where the spin inner product occurs instead of the unit
operator in spin space.

\begin{figure}
\begin{center}
\vskip 2cm
\includegraphics[width=11.5cm,angle=0]{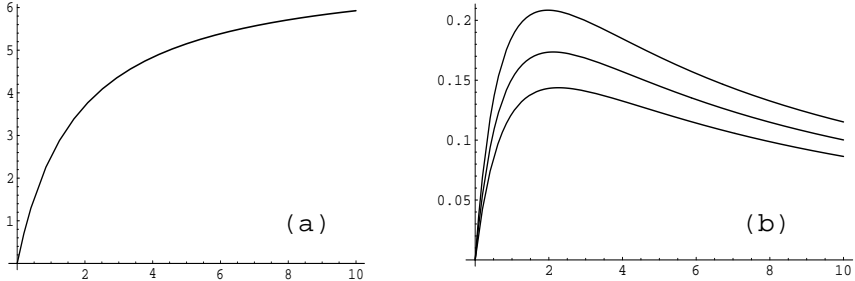}
\end{center}
\vskip -0.4cm
\caption{
Interaction constants (in units of $J=t^4/\Delta^3$) as a function of
$U_p/\Delta$ for $J_p/U_p=0.1$ and $J_{\rm H}/\Delta=0.05$:
(a) Orbital interaction constant $J_{\tau}$;
(b) Spin-orbital interaction constants $J_{\mu}$, $J_{\nu}$ and
$J_{\sigma}$ (from top to bottom).
}
\label{fig:IntconUp}
\end{figure}

To get an idea of the order of magnitude of the SE coupling constants
one would need to estimate the parameters involved. Unfortunately,
reliable estimates are available almost exclusively for the divalent
TM ions, while only little work has been done on the trivalent ions
\cite{Zaa90,Boc92,Sai95}. Based upon what is available we estimate
(all values in eV): $t \simeq 1.5$, $J_{\rm H} \simeq 1.1$,
$U_p \simeq 5$, $J_p \simeq 0.8$ \cite{Gra92} and $\Delta \sim 2$,
where the last value is the most uncertain. The small value of $\Delta$
suggests that the covalency effects are quite important in LiNiO$_2$,
and thus the present model has to
be considered only as describing the generic structure of the
interactions between states with spin $S=1/2$ and $e_g$-type orbital
degeneracy, which originate from Ni$^{2+}$ ions surrounded by a hole
shared between the Ni $3d$ orbitals and the oxygen $2p$ orbitals.
It is anyway obvious that $\Delta$ is too small to consider the
perturbation theory a controlled expansion. We shall therefore regard
the overall energy scale $J \equiv t^4 / \Delta^3$ as an unknown
parameter and take it as our energy unit. We can then study the
relative size of the coupling constants, which are controlled by
the ratios $U_p/\Delta$, $J_{\rm H}/\Delta$ and $J_p/U_p$, which we
treat as parameters. Where any of them needed to be fixed for varying
other variables, we took
$U_p/\Delta=2$, $J_{\rm H}/\Delta=0.05$ and $J_p/U_p=0.1$
(which is maybe a bit small for the first and especially the second
and third one).

The variation of the orbital SE coupling constant $J_{\tau}$ and the
three spin and spin-orbital SE constants $J_{\sigma}$, $J_{\nu}$ and
$J_{\mu}$ with
the strength of the oxygen Coulomb repulsion $U_p$ is shown in figure
\ref{fig:IntconUp}. It illustrates once again that oxygen Coulomb
interaction is crucial: at $U_p=0$ all coupling constants vanish.
One further notes that the inequality (\ref{Jinequality}) is well
satisfied: indeed $J_{\tau}$ is by far the largest, and the three
spin coupling constants are in the expected order. Note also that
there is more or less saturation for $U_p/\Delta$ larger than $\simeq 2$
so that the inaccuracy in this parameter is not so important.

\begin{figure}
\begin{center}
\vskip 2cm
\includegraphics[width=11.5cm,angle=0]{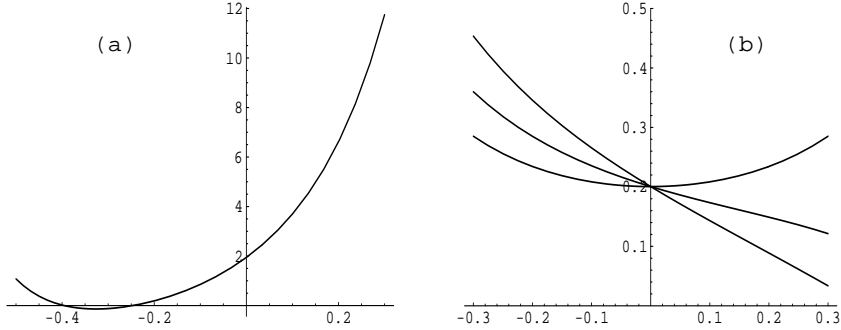}
\end{center}
\vskip -0.4cm
\caption{
Interaction constants (in units of $J=t^4/\Delta^3$) as a function of
$2 J_{\rm H}/\Delta$ for $U_p/\Delta=2.0$ and $J_p/U_p=0.1$:
(a) Orbital interaction constant $J_{\tau}$;
(b) Spin-orbital interaction constants $J_{\mu}$, $J_{\nu}$ and
$J_{\sigma}$ (from top to bottom for positive $J_{\rm H}$).
}
\label{fig:IntconJH}
\end{figure}

Figure \ref{fig:IntconJH} shows the variation with $J_{\rm H}$, the
strength of the Hund's rule exchange on nickel. One observes that a
finite $J_{\rm H}$ enhances the orbital SE considerably
[figure \ref{fig:IntconJH}(a)], basically because this lowers the energy
of the triplet and so further disfavours (see the remark at the end of
section \ref{sec:SEeg}) the N configuration, already disfavoured because
it allows the largest number of hopping sequences.
Moreover, figure \ref{fig:IntconJH}(b) shows explicitly that nonzero
$J_{\rm H}$ is essential for having unequal spin interaction constants.
Since this is equivalent to $J_{\rm M}^S$ being nonzero (and positive)
and thus for having potentially AF interaction, as argued above when the
Hamiltonian (\ref{Ham1}) was discussed, inclusion of the Ni$^{2+}$ term
splitting makes a {\it qualitative} difference for the obtained
spin-orbital model. It is amusing that in the unphysical regime of
negative $J_{\rm H}$ the order of the spin SE constants is inverted, so
that $J_{\rm M}^S$ would be negative and AF interactions would not be
possible.
Note further that magnetic frustration is apparently maximal at 
$J_{\rm H}=0$,
where the three types of spin dependent interaction are equally strong
and so experience the strongest mutual competition.


\section{Interaction between a pair of Ni$^{3+}$ ions}
\label{sec:pair}

\begin{figure}
\begin{center}
\vskip 2cm
\includegraphics[width=11.5cm,angle=0]{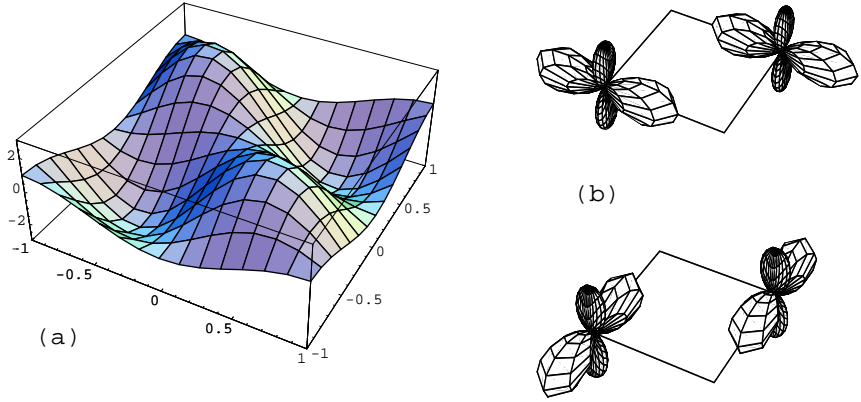}
\end{center}
\vskip -0.4cm
\caption{
Orbital superexchange $J_{\tau}$ for a single pair along $XY$ (see
figure \protect\ref{fig:structure}):
(a) bond energy as a function of the orbital angles $\theta_1$ and
$\theta_2$
(energy in units of $\frac{1}{4}J_{\tau}$, angles in units of $\pi$);
(b) the two favoured pairs of occupied $e_g$ orbitals.
}
\label{fig:pairorbs}
\end{figure}

In view of the dominating size of the orbital interaction $J_{\tau}$ it
is natural to consider first the orbital interactions alone. Before
addressing the Hamiltonian (\ref{HamT}) on the entire triangular plane,
it is useful to solve first the problem of a single Ni$^{3+}$--Ni$^{3+}$
pair, which for definiteness we assume to have its bond in the $XY$
plane. For an arbitrary orbital pair state
$|i \theta_i \rangle |j \theta_j \rangle$ the orbital
energy of the pair is then given by [compare equation (\ref{bondexpW})]
\begin{equation}
\label{pairEnorb}
  E_{T}^{\rm pair}(\theta_i, \theta_j) =
\case{1}{4} J_{\tau} [ 4 \cos^2 (\case{\theta_{i}+\theta_{j}}{2})
      - 2 \cos^2 (\case{\theta_{i}-\theta_{j}}{2}) - 1 ] ,
\end{equation}
and the corresponding energy surface is shown in figure
\ref{fig:pairorbs}(a). One notes that similar orbitals
($\theta_i \simeq \theta_j$) are generally favoured over dissimilar
orbitals ($\theta_i \simeq \theta_j \pm \pi$).
More in particular, for a single bond ${\cal H}_{\rm eff, o}$ is seen
to favour {\it specific\/} identical orbitals, viz. those with
$\theta_i = \theta_j = \pm \case{\pi}{2}$, i.e. a pair configuration
with orbitals
$( |z \rangle \pm |\bar{z} \rangle ) / \sqrt{2}
\propto d_{3z^2-r^2} \pm d_{x^2-y^2}
$
at both sites, as illustrated in figure \ref{fig:pairorbs}(b),
with energy $E_{T}^{\rm pair}=-\case{3}{4} J_{\tau}$.
%
%
%
Frustration prevents this optimum orbital arrangement to be realized on
all bonds simultaneously, as already pointed out by MK.
As will be seen in the next section, on the
full triangular lattice the orbital interaction favours
phases in which a pair has either a ferro orbital (FO) arrangement at a
general angle $\theta \equiv \theta_i = \theta_j$ or similarly a
canted orbital (CO) arrangement with $\theta\equiv \theta_i =-\theta_j$.
It is therefore of interest to consider the effective spin-spin
interactions realized under those conditions.

Quite generally, once the orbitals of the pair are fixed, supposedly
determined by the pure orbital interaction, the resulting effective
spin SE interaction can be written as a Heisenberg Hamiltonian,
\begin{equation}
\label{Heis}
{\mathcal H}_{\rm spin}^{\rm eff} (\theta_i, \theta_j) =
 \sum_{\langle ij \rangle}  J_{ij}^{\rm eff}(\theta_i, \theta_j)
    \: {\bi S}_i \cdot {\bi S}_j ,
\end{equation}
with the effective spin SE coupling $J_{ij}^{\rm eff}$ given by
\begin{equation}
\label{Jeff}
\fl J_{ij}^{\rm eff}(\theta_i, \theta_j) = - \case{1}{4}
   \Big[ 4 J_{\sigma}
         - J_{\mu} \Big( 1 + 2 \cos^2 \theta_{-}
                   - 4 \cos^2 (\theta_{+} + \chi_{\gamma}) \Big)
       - 4 J_{\nu} \cos (\theta_{+} + \chi_{\gamma}) \cos \theta_-
   \Big],
\end{equation}
where $\theta_{\pm}=(\theta_i \pm  \theta_j)/2$, as obtained by
inserting equations (\ref{bondexpV}) and (\ref{bondexpW}) into
(\ref{HamTS}).
For the two relevant cases of orbital order one thus finds, by adopting
the appropriate values
for the angles $\theta_i$ and $\theta_j$,
\begin{eqnarray}
\label{JspinFO}
\fl J_{\rm FO}(\theta) \equiv J_{ij}^{\rm eff}(\theta, \theta) =
  - \case{1}{4} \Big[ 4 J_{\sigma} - 3 J_{\mu}
                    + 4 J_{\mu} \cos^2 (\theta +\chi_{\gamma})
                    - 4 J_{\nu} \cos (\theta + \chi_{\gamma}) \Big], \\
\label{JspinCO}
\fl J_{\rm CO}(\theta) \equiv J_{ij}^{\rm eff}(\theta,-\theta) =
  - \case{1}{4} \Big[ 4 J_{\sigma} -(1-4\cos^2 \chi_{\gamma})J_{\mu}
                    - 2 J_{\mu} \cos^2 \theta
                    - 4 \cos \chi_{\gamma} J_{\nu} \cos \theta
                \Big] .
\end{eqnarray}

\begin{figure}
\begin{center}
\vskip 2cm
\includegraphics[width=11.5cm,angle=0]{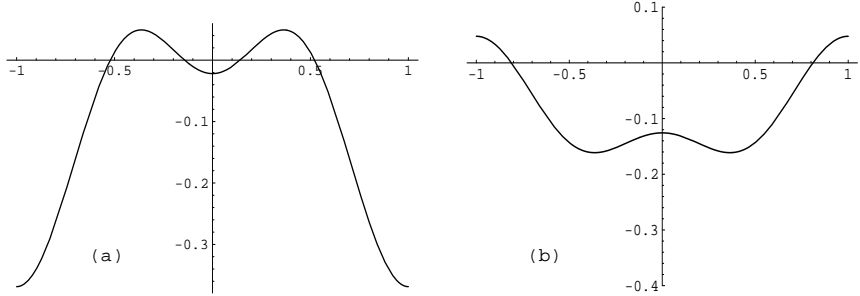}
\end{center}
\vskip -0.4cm
\caption{
Effective spin-spin superexchange constant (in units of $J$) versus
angle $\theta$ (in units of $\pi$; see equation (\ref{orbital})):
(a) For a single pair along $XY$ for ferro orbitals
($\theta=\theta_i=\theta_j$);
(b) For a single pair along $YZ$ or $ZX$ for canted orbitals
($\theta=\theta_j=-\theta_j$).
}
\label{fig:spinspin}
\end{figure}

The dependence on $\theta$ is shown in figure \ref{fig:spinspin} for the
values of $J_{\sigma}$, $J_{\mu}$ and $J_{\nu}$ obtained with our
standard parameter set.
Figure \ref{fig:spinspin}(a) shows that in the FO case the
spin SE constant is mostly FM (negative), as expected from the
Goodenough-Kanamori-Anderson rules, but that AF values, though smaller,
are indeed possible, notably for orbitals with
$\theta \simeq \pm \pi/3$. Note that the coupling constant is that for a
bond along $XY$; equation (\ref{JspinFO}) shows that the curve should be
shifted by $\pm 2\pi/3$ for bonds along $YZ$ or $ZX$.
The CO case is shown in figure \ref{fig:spinspin}(b),
for a bond in the $YZ$ or $ZX$ direction while the angle is still given
with respect to the basis
\{$|i z\rangle$, $|i \bar{z}\rangle$\} as in equations (\ref{basis}) and
(\ref{orbital}). Note that the coupling can again be AF as well,
precisely for those angles ($\theta \simeq \pm \pi$) where the FO
arrangement gives the largest FM coupling in the $XY$ direction.

\section{Possible ground states}
\label{sec:GS}


We have already noticed that the orbital SE is
{\it strongly frustrated\/} --- since the specific orbitals minimizing
the pair energy are different for each bond direction, the interactions
between neighboring Ni ions cannot be satisfied in all three bond
directions simultaneously. The fact that frustration occurs although the
interaction is of ferro type is somewhat unusual and is generic for
orbital physics as compared to spin physics. The underlying reason is
that the orbital interaction Hamiltonian (\ref{HamT}) does not have
global continuous rotational $U(1)$ symmetry with regard to the orbital
angles $\theta_i$, but is only invariant under rotation by an angle
$\pm 2 \pi/3$, corresponding to a cyclic permutation of the bond
directions $XY$, $YZ$ and $ZX$. This characteristic feature and the
ensuing frustration are not captured by $SU(4)$ symmetric models
\cite{Li98,Pen03}.

Next we consider what states with long-range order (LRO) are possible.
As pointed out by MK, the simplest such state, viz. with uniform
ferro orbital order, is actually a minimum energy state in MF
theory. Its energy is independent of $\theta$ and equal to
$-\case{3}{4} J_{\tau}$ per site, while moreover this FO phase may be
interrupted without energy cost by lines in which all orbitals have
the opposite angle $-\theta$, i.e. by canting of all orbitals along
a line in the triangular lattice.
In order to see how this comes about, one may write down the energy for
a set of FO lines (for definiteness along the $XY$ direction) with
orbitals in line $n$ given by $\theta_n$. From (\ref{HamT}) and
(\ref{bondexpW}) one obtains
\numparts
\begin{eqnarray}
\fl E_{\rm FO-l} = E_{\rm intraline} +  E_{\rm interline}
                                \nonumber \\
\lo = \case{1}{4} J_{\tau} \sum_n  [ 2 \cos 2 \theta_{n} - 1 ]
                                \nonumber \\
  + \case{1}{2} J_{\tau} \sum_{n}   \Big[
       \cos \, (\theta_{n} + \theta_{n+1} - \case{2 \pi}{3})
     + \cos \, (\theta_{n} + \theta_{n+1} + \case{2 \pi}{3})
     - \cos \, (\theta_{n} - \theta_{n+1} )        \Big]
                                \nonumber \\
\label{interline:a}
\lo = \case{1}{4} J_{\tau} \sum_n  [ 4 \cos^2 \theta_{n} - 3 ]
 - J_{\tau} \sum_{n}
      \cos \theta_{n} \;  \cos \theta_{n+1}
\\
\label{interline:b}
\lo = \case{1}{4} J_{\tau} \sum_{n}
   \Big[ 2 ( \cos \theta_{n} - \cos \theta_{n+1} )^2 - 3 \Big] .
\label{interline}
\end{eqnarray}
\endnumparts
Equation (\ref{interline:a}) shows that upon variation of the angles
$\theta_n$ the gain in intraline line energy is compensated by a loss in
interline energy, leading on balance to $E_{\rm FO-l}$ being not
dependent on the value of $\theta$ in the ordered phase, while
equation (\ref{interline:b}) shows that for any two neighbouring lines
the same energy is obtained for
$\theta_n=+\theta_{n+1}$ and $\theta_n=-\theta_{n+1}$.
So in addition to the degeneracy with respect to $\theta$ generated by
the frustration, the model also gives rise to randomness due to
uncorrelated switches from $+\theta$ to $-\theta$ between successive
lines. Such a random CO phase is probably best interpreted as the
occurrence of twin boundaries or antiphase boundaries between equivalent
FO domains. Because of the zero energy cost for formation of a boundary
there is actually no preference for the formation of large domains.

It was argued by MK that the degeneracy in $\theta$ is an artefact of
MF theory, and that this $U(1)$ symmetry of the FO state, not present in
the original Hamiltonian (\ref{HamT}), would be lifted in the FO domains
by the `order-out-of-disorder' mechanism \cite{Vil80}, and that
in this way the threefold symmetry associated with the triangular
lattice would be restored.
However, as the energies involved in the restoring quantum fluctuations
are generally small, only a fraction of the orbital coupling constant
$J_{\tau}$, one should at this stage reconsider the orbital-spin
coupling. Even though the relevant coupling constants $J_{\sigma}$,
$J_{\mu}$ and $J_{\nu}$ are much smaller than $J_{\tau}$, the
spin-and-orbital dependent SE energy might still be more significant
than the fluctuation energies in lifting the $\theta$ degeneracy.
Figure \ref{fig:spinspin} suggests that it could do so already at the
MF level and so determine the ground state before fluctuations need to
be considered.

\begin{figure}
\begin{center}
\vskip 1cm
\includegraphics[width=15.5cm,angle=0]{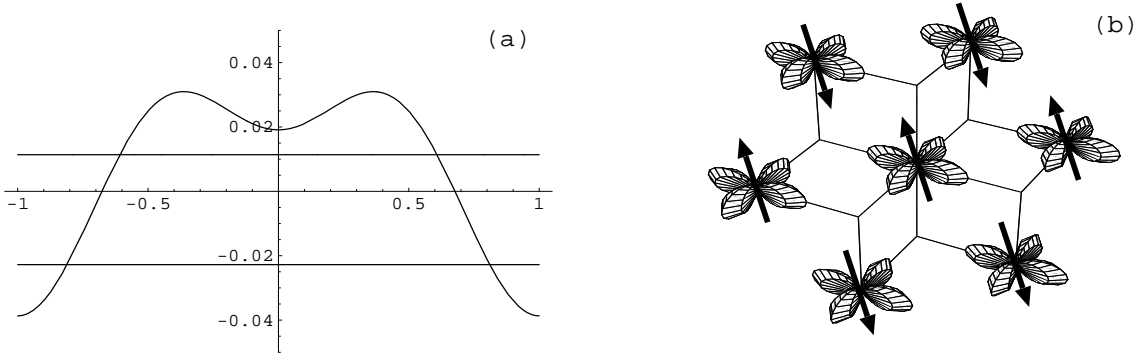}
\end{center}
\vskip -0.4cm
\caption{
Spin-and-orbital order:
(a) MF energy per site (in units of $J=t^4/\Delta^3$) for
various ordered phases as function of the orbital angle $\theta$:
(upper horizontal line:) ferro-orbital / 120$^{\circ}$ spin order ;
(lower horizontal line:) ferro-orbital / ferromagnetic order, degenerate
with orbital-line / ferromagnetic order;
(curve:) ferro-orbital/ spin-line order, degenerate with orbital-line
/ spin-line order;
(b) Ground state in mean field:  ferro-orbital order of $d_{x^2-y^2}$
orbitals with spin-line order (along the $XY$ direction).
}
\label{fig:spinorbMF}
\end{figure}

%
%
We therefore consider the following LRO spin patterns in conjunction
with orbital order: (i) FM order; (ii) the 120$^{\circ}$
three-sublattice order which is the ground state of the classical TALAF;
(iii) AF line order similar to that of the orbitals, i.e. lines of
parallel spins along the same direction as the orbital lines, and the
spins alternating between up and down on successive lines. The result
for the MF energy is shown in figure \ref{fig:spinorbMF}(a).
It turns out that indeed the $\theta$ degeneracy is lifted, and the
lowest MF energy is seen to be obtained for FO order with
orbital angle $\theta=+\pi$ (or $\theta=-\pi$), corresponding to a
$d_{x^2-y^2}$ orbital at every site, in combination with spin AF line
order, illustrated in figure \ref{fig:spinorbMF}(b).
This shows that even though the spin-orbital interaction is considerably
weaker than the pure orbital interaction, it can still have a
qualitative effect on the nature of the ground state by breaking the
$U(1)$ symmetry generated by the orbital interactions alone and restoring
threefold symmetry. Therefore the large difference in coupling strength
does not necessarily lead to a decoupling of the spin and orbital
degrees of freedom.

Remarkably, the $+\theta$/$-\theta$ twin-boundary degeneracy persists,
not only at the optimum angle $\pm \pi$, where it is immaterial since it
corresponds only to a sign change of the orbital wavefunctions, but also
at arbitrary $\theta$. This could in fact have been recognized from the
$\theta$ dependence of the effective spin exchange constants: the spin
exchange in the transverse direction (on the bonds along $YZ$ and $ZX$)
as given for the FO case by shifts over $\pm \case{2\pi}{3}$ in figure
\ref{fig:spinspin}(a) is the same as that given for the CO case in
figure \ref{fig:spinspin}(b).

Most significantly, we note that both for the ordered FO phase and for
the disordered CO phase the effective spin-spin exchange is
\textit{antiferromagnetic\/} in the transverse direction, also for
orbital angles deviating somewhat from the optimum value $\pm \pi$.
However, for a disordered phase with truly random orbital angles, either
a paraorbital state at high temperature or an orbital glass, the
situation would be very different. Figure \ref{fig:spinspin} indicates
that orbital randomness would produce a wide distribution of mostly FM
exchange constants, although small AF values would still occur but not
in a regular pattern as in the FO and CO phases. Obviously a phase
transition to an orbital ordered phase would manifest itself also in
the magnetic properties, e.g. in the dispersion of spin waves.

\begin{figure}
\begin{center}
\vskip 0.5cm
\includegraphics[width=11.5cm,angle=0]{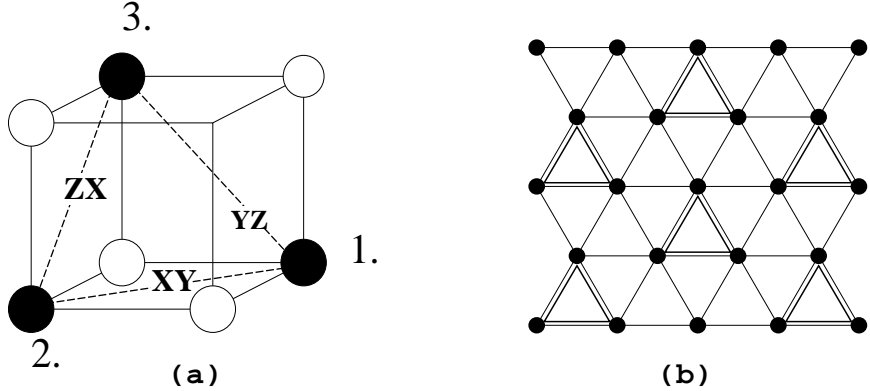}
\end{center}
\vskip -0.4cm
\caption{
Geometry relevant for the orbital valence bond (OVB) state:
(a) Triangular three-site plaquette in a Ni plane for determination of
the exact eigenstates of the orbital interaction;
(b) Lattice covering with three-site plaquettes in the OVB state.
}
\label{fig:OVB}
\end{figure}

%
%
Whereas the above analysis shows that the spin-orbital interaction could
support orbital LRO, the degeneracy in the orbital sector and the small
MF energy of only $-\case{1}{4} J_{\tau}$ per bond, both for the
FO and for the CO phase, raise the question whether orbital LRO by
itself is stable against formation of a disorded state of valence bond
(VB) type. We therefore studied orbital valence bond (OVB) states
\cite{Fei97,Li98} in the following way.
First we solved the eigenvalue problem of the orbital SE Hamiltonian
${\mathcal H}_{\rm eff,o}$ exactly on the three-site triangular
plaquette shown in figure \ref{fig:OVB}(a). The eigenstates are four
doublets, with ground state energy $E_0^{\rm plaq}=-\case{3}{2}J_{\tau}$
and excited states at $-J_{\tau}$, $+\case{1}{2}J_{\tau}$ and
$+2J_{\tau}$, the ground state doublet being
\begin{eqnarray}
\vert e_1\rangle = \sqrt{\case{4}{7}}
   (|x\rangle_1 |z\rangle_2 |y\rangle_3
   +|z\rangle_1 |y\rangle_2 |x\rangle_3) ,
                                \nonumber \\
\vert e_2\rangle = \sqrt{\case{4}{7}}
  (|\bar{x}\rangle_1 |\bar{z}\rangle_2 |\bar{y}\rangle_3
+  |\bar{z}\rangle_1 |\bar{y}\rangle_2 |\bar{x}\rangle_3 ) .
\label{OVBbasis}
\end{eqnarray}
Note that particular combinations of orbitals are favoured, similar
to the orbital order for the V triangle in LiVO$_2$ \cite{Pen97}.
It is remarkable that on a triangular lattice such triangular VB
correlations in orbital space are favoured, in contrast to dimer singlet
correlations in spin systems \cite{Moe01}.

We then constructed an OVB solid by covering the triangular lattice with
three-site plaquettes, as in figure \ref{fig:OVB}(b), and assigning to
each triangular plaquette $l$ an orbital wavefunction built from the
groundstate components (\ref{OVBbasis}),
\begin{equation}
\vert \Delta_l\rangle =
                \cos \theta_l \, \vert e_1\rangle_l
              + \sin \theta_l \, \vert e_2\rangle_l ,
\label{Delta}
\end{equation}
with $\theta_l$ still to be determined. This amounts to assigning a
fictitious spin to each plaquette and solving a frustrated spin problem
on the triangular lattice. In MF theory the total energy is now
given by addition of the groundstate energies from all plaquettes and
the interplaquette energies coming from the two bonds between nearest
neighbour plaquettes,
\begin{equation}
\fl E_{\rm OVB} = \sum_{l} E_{\Delta_l}
      + \sum_{l,m} \langle \Delta_l, \Delta_m|{\mathcal H}_{\rm eff,o}
                          | \Delta_l, \Delta_m \rangle
= \sum_{l} E_0^{\rm plaq}
  + \sum_{l,m} E_{\rm VB}^{\rm p-p}(\theta_l, \theta_m) .
\label{E-OVB}
\end{equation}
For a uniform solution the energy per bond is given by
\begin{equation}
\bar{E}_{\rm OVB} =  \case {1}{6}
     \Big( 2 \times \case{1}{3} \times E_0^{\rm plaq}
         + 4 \times \case{1}{2} \times E_{\rm VB}^{\rm p-p} \Big),
\label{OVBbond}
\end{equation}
by counting the number of intraplaquette and interplaquette bonds.
Although the plaquette eigenenergy amounts to
$\case{1}{3} E_0^{\rm plaq} = -\case{1}{2}J_{\tau}$
per bond of the plaquette, which for this exact solution is of course
lower than the energy per bond for the FO and CO phases, the plaquettes
cover only one third of the bonds of the lattice, and so the
contribution from the interplaquette energy $E_{\rm VB}^{\rm p-p}$ is
decisive.

The optimum values of the plaquette angles are found to be
$\theta_l=+\case{\pi}{4}$, $\theta_m=-\case{\pi}{4}$, for which
$E_{\rm VB}^{\rm p-p}=-\case{27}{49} J_{\tau}$, and if this value could
be attained between any pair of plaquettes one would have
$ \bar{E}_{\rm OVB} = - ( \case {1}{6} +\case{9}{49}) J_{\tau}
  \simeq -0.3503 J_{\tau} $,
lower than $\bar{E}_{\rm FO}=\bar{E}_{\rm CO}=-0.25 J_{\tau}$.
However, this would requiring alternating angles $\pm \case{\pi}{4}$,
which is impossible because of frustration on the triangular lattice.
The optimum value that can be obtained is
$\bar{E}_{\rm OVB} \simeq -0.228 J_{\tau}$, realized for instance
by line order of $\pm \case{\pi}{4}$. This is remarkably close to the
energy per bond of the site-ordered states,
and just confirms that the orbital sector is strongly frustrated.
It is therefore likely that some quenched disorder would suffice to
turn the system into an orbital glass below a freezing temperature
$T_{\rm of}$ determined by $J_{\tau}$, as proposed to explain the
behaviour of the orbitals observed in LiNiO$_2$ \cite{Rey01}.

\section{Origin of the difference between LiNiO$_2$ and NaNiO$_2$}
\label{sec:compounds}

In view of the above results the properties of LiNiO$_2$ are puzzling
\cite{Nun00,Rey01}, as neither orbital nor spin order sets in down to
the lowest temperatures. Thus the high degeneracy of the ground state
manifold apparently persists, with  Ni$^{3+}$ ions being in the low-spin
($t_{2g}^6e_g^1$) state, with twofold orbital and twofold spin
degeneracy. The standard scenario for such ions would be a cooperative
Jahn-Teller effect lifting the orbital degeneracy below a structural
transition at $T_s$, followed by a magnetic transition at $T_N$, lifting
also the degeneracy in spin space. Both transitions could be induced
already by the electronic mechanism involved in the SE, but the
interactions with the lattice are expected to play a significant role in
the structural transition, enhancing the value of $T_s$,
as found in the manganites \cite{Fei99}.
In spite of the strong orbital interactions reported above, and although
the $S=1/2$ spins interact by the predominantly FM SE
described in the previous sections, and the magnetic susceptibility
follows the Curie-Weiss behaviour,
$\chi\propto (T-\theta_{\rm CW})^{-1}$ with $\theta_{\rm CW}\sim 35$ K
down to about 80 K \cite{Kem90,Hir91,Yam96}, long-range order in either
orbital or spin space is absent \cite{Kit98,Nun00}. This would suggest
that the $e_g$ orbital degree of freedom plays an important role,
possibly stabilizing strong singlet orbital correlations on individual
bonds, and supporting a {\it spin liquid\/} state \cite{Fei97}.

Before addressing the central question of the origin of the spin liquid
state in LiNiO$_2$, let us summarize shortly the structural and magnetic
properties of the similar NaNiO$_2$ compound, which crystallizes in the
same structure as LiNiO$_2$ (see figure \ref{fig:structure}).
The standard scenario with two subsequent transitions described above is
indeed realized in NaNiO$_2$. The latter compound undergoes a
first-order cooperative Jahn-Teller transition lowering the local
symmetry from trigonal to monoclinic at $T_s=480$ K \cite{Chb00}, and the
Ni-O distances change from $d=1.98$ \AA{} to two long bonds with $d=2.14$
\AA{} and four short bonds with $d=1.91$ \AA{} at $T<T_s$. In the
low-temperature phase a magnetic transition at $T_N=20$ K to an $A$-type
AF insulator follows --- it was first derived from magnetization
measurements on a single crystal by Bongers and Enz already in 1966
\cite{Bon66}. This magnetic phase has also been confirmed by a
complete static and dynamic magnetic study at low temperatures, where
strong anisotropy between the effective magnetic interactions was
established \cite{Chb00}. An FM intralayer coupling of
$J_{\rm FM}=-26$ K and a weak AF interlayer coupling of $J_{\rm AF}=2$ K
were found, values consistent with the observed Curie-Weiss temperature
$\Theta=35$ K and N\'eel temperature $T_N=20$ K. Similar values of the
exchange constants ($J_{\rm FM}\simeq -29$ K and
$J_{\rm AF}\simeq 1.9$ K) were also deduced recently by analyzing the
dispersion of spin waves found in neutron scattering experiments
\cite{Lew04}. We remark that the parameters which fit the Curie-Weiss
law change around the structural transition at $T_s$ \cite{Chb00}, which
indicates that the magnetic couplings depend on the actual orbital
state, in agreement with the model presented here.

In contrast, in LiNiO$_2$ the structural transition is absent, but EXAFS
experiments indicate the presence of local Jahn-Teller distortions
below an orbital freezing temperature $T_{\rm of}\sim 400$ K, with two
different Ni-O distances again favouring occupied directional
$d_{3z^2-r^2}$-type orbitals \cite{Rou95}:
two long bonds with $d=2.09$ \AA{} and four short bonds with $d=1.91$
\AA{}. While these local distortions are similar to those observed
in NaNiO$_2$, the absence of a macroscopic distortion in LiNiO$_2$ has
been a mystery since its synthesis \cite{Bon57}. However,
very recent evidence from neutron diffraction indicates that below
$T_{\rm of}$ the orbitals actually develop short-to-medium-range order
in a trimerized state, although the associated strain field prevents the
ordering from becoming long-range, and instead nanoscale domains of
local orbital trimer order are formed \cite{Chu05}. Nevertheless,
even local
distortions favouring occupied $d_{3z^2-r^2}$ orbitals are intriguing,
as we have seen above that the SE favours instead alternating symmetric
and antisymmetric combinations of the basis orbitals,
$d_{3z^2-r^2}\pm d_{x^2-y^2}$, for individual Ni$^{3+}$-Ni$^{3+}$
pairs (so e.g. in a disordered state), and FO order
of $d_{x^2-y^2}$ orbitals, as obtained in the MF approximation.
Actually, such an ordered phase with occupied
$d_{x^2-y^2}$ orbitals has been observed in large magnetic
field \cite{Bar99}, which demonstrates that the state predicted by the
SE model is energetically close to the actual ground state, and
a quantum phase transition between different orbital states
can be triggered by an applied magnetic field.
The very fact that instead, in the absence of a magnetic field,
local distortions of NiO$_6$ octahedra associated with occupied
$d_{x^2-y^2}$-type orbitals dominate at low temperature, we interpret as
a strong indication that the orbital physics is in the end determined by
the Jahn-Teller coupling to the lattice and not by the orbital SE.

The ground state of LiNiO$_2$ remains the subject of intense debate.
Several mechanisms for the spin liquid phase have been proposed so far:
($i$) quantum fluctuations on AF bonds \cite{Fei97},
($ii$) geometrical frustration of AF interactions in the $SU(4)$
spin-orbital model on the triangular lattice \cite{Li98,Pen03},
($iii$) decoupling of spin and orbital degrees of freedom
because of the large difference, confirmed above, between orbital and
spin-orbital SE coupling constants \cite{Mos02},
($iv$) disorder due to Ni$^{2+}$ ions in the Li planes
\cite{Bar99,Nun00}.
Interestingly, also in the $SU(4)$ model both AF and FM interactions are
allowed \cite{Li98,Pen03}, so at first sight the mechanism ($ii$) might
appear similar to the present model of section \ref{sec:model}. However,
the $e_g$ orbital operators do not respect $SU(2)$ symmetry, and thus the
actual magnetic interactions which follow from the $SU(4)$ model are
quite different. For instance, in the $SU(4)$ model one may assume
perfect FO order, with
$\langle {\bi T}_i \cdot {\bi T}_j\rangle=\frac{1}{4}$ for the orbital
operators on each bond $\langle ij\rangle$, which would give AF spin
interactions on all the bonds, while such a situation cannot be realized
for $e_g$ orbitals. Also a state with the orbitals staggered for all
bonds is not allowed on a triangular lattice. Therefore, we suggest
that the role of geometrical frustration of magnetic interactions is
overestimated in the $SU(4)$ model and cannot explain the properties of
the triangular Ni planes by itself. In contrast, all other mechanisms
could contribute and stabilize the spin liquid state in a 3D lattice of
LiNiO$_2$. We propose a new scenario below in which the above aspects
(iii) and (iv) are supplemented by two important features of the orbital
%
glass \cite{Rey01} or trimer nanodomain state \cite{Chu05} below the
transition at $T_{\rm of}\simeq 400$ K:
($v$) the weakness of the (effective) SE interactions between Ni$^{3+}$
ions in adjacent Ni planes, and
($vi$) the Jahn-Teller coupling to the lattice, which induces local
distortions favouring directional $d_{3z^2-r^2}$-like orbitals.

We emphasize that the experimental studies on NaNiO$_2$ and LiNiO$_2$
have identified two competing effects which operate in triangular 2D Ni
planes:
on the one hand local Jahn-Teller distortions favouring occupied
$d_{3z^2-r^2}$ orbitals, produced by the on-site Jahn-Teller coupling,
and on the other hand the SE, which
would instead stabilize orbital order with occupied $d_{x^2-y^2}$
orbitals, coexisting with FM interactions within the planes.
Although Ni-Ni distances and Ni-O-Ni bond angles in the O--Ni--O
trilayers of the two compounds are similar, it may well be that the
in-plane elastic couplings and orbital SE differ just sufficiently that
the 2D orbital susceptibilities are qualitatively different.
In particular, the competition between Jahn-Teller effect and SE may
%
favour orbital disorder or complex multiple-sublattice orbital order
when one interaction is only marginally stronger
than the other one.
This is in fact suggested by the behaviour of LiNiO$_2$ in magnetic
field, mentioned above, and by the properties of the mixed compound
Li$_{0.3}$Na$_{0.7}$NiO$_2$ \cite{Hol04}, discussed below. Whether
orbital and magnetic order stabilize in a 3D system, however, depends
also crucially on the interplane interactions. Without any coupling
between the planes, no long-range order could occur in the 2D planes
anyway, in accordance with the Mermin-Wagner theorem.

One thus expects that 3D {\it orbital order\/} can be induced only by a
sufficiently strong effective interplane coupling, as apparently
encountered in NaNiO$_2$, while the absence of orbital order observed
in Li$_{1-x}$Ni$_{1+x}$O$_2$ may be explained by a weaker effective
interplane coupling in combination with an already weaker in-plane
ordering tendency. Even without an explicit extension of the present
microscopic spin-orbital model we can already suggest two mechanisms
responsible for this weak interlayer coupling. First of all, the
interplane SE is intrinsically weaker in LiNiO$_2$ than in NaNiO$_2$
because of the different covalency of the Ni--O--Li--O--Ni bonds as
compared to the Ni--O--Na--O--Ni bonds,
owing to the weaker hybridization between the oxygen $2p$ orbitals and
the spatially less extended Li $2s$ orbitals
than between the oxygen $2p$ orbitals and the bigger Na $3s$ orbitals.
Although this difference is quantitative, it
may lead to a qualitatively different behaviour.
A further point is then how the individual Ni$^{3+}$-Ni$^{3+}$
orbital interactions between two neighbouring planes add up.
For two orbital ordered planes, as in NaNiO$_2$, the interactions add up
coherently, but if the orbitals preferentially avoid ordering in each Ni
plane, e.g. because of a tendency towards domain formation,
then in spite of the intrinsically identical spin-orbital interaction
between Ni$^{3+}$ ions in adjacent Ni planes, orbital (and magnetic)
couplings add up incoherently.
Second, it is well known that in the case of NaNiO$_2$ an almost ideal
structure is obtained, while Li$_{1-x}$Ni$_{1+x}$O$_2$ is never
stoichiometric and extra Ni ions occur in Li planes \cite{Nun00}.
This creates frustration due to extrinsic randomness in the individual
interplane interactions, and thus further weakens the effective
interplane coupling.

Finally, the absence of 3D {\it magnetic order\/} in LiNiO$_2$ is
apparently similarly due to the weakness of the magnetic effective
interplane coupling.
While it is tempting to ascribe it instead to the presence of both AF
and FM SE interactions in the triangular Ni planes, demonstrated above
in the spin-orbital model, this is unlikely to be the leading effect
stabilizing the spin liquid. In particular, in the orbital glass state
the SE would be FM on by far the majority of the bonds in the Ni planes,
as follows from equation (\ref{Jeff}) [compare also figure
\ref{fig:spinspin}], with the precise fraction depending on the degree
of disorder, while moreover the few AF interactions would be much weaker
than the FM interactions.
That the effective interplane magnetic coupling in LiNiO$_2$ is weak, is
not only due to the interplane SE being intrinsically weak owing to the
small covalency.
In addition, in spite of the identical spin-orbital interaction between
%
Ni$^{3+}$ ions in adjacent Ni planes, the absence of orbital long-range
order in
the Ni planes will make individual Ni$^{3+}$-Ni$^{3+}$ spin-spin SE
interactions add up incoherently.

An additional extrinsic mechanism opposing magnetic order is provided
by magnetic interactions due to the Ni$^{2+}$ defects in the Li planes,
which by themselves frustrate weak AF SE interactions by stronger FM
interactions along Ni$^{3+}$-Ni$^{2+}$-Ni$^{3+}$ units
\cite{Nun00,Cha02}, additionally enhanced due to the $S=1$ spin states
involved. As long as such defects do not occur, coherence in the
magnetic interplane coupling is easier to obtain. Our scenario is not
inconsistent with the recent observation of the decoupling of orbital
and spin degrees of freedom in the mixed compound
Li$_{0.3}$Na$_{0.7}$NiO$_2$ which has a very different orbital state
and yet a similar magnetic ground state as NaNiO$_2$ \cite{Cha00},
with similar exchange constants \cite{Hol04}.

\section{Discussion and summary}
\label{sec:summ}

The above analysis demonstrates that the qualitative properties which
follow from orbital and spin correlations within triangular Ni planes
with 90$^{\circ}$ Ni--O--Ni bonds can be understood within a realistic
spin-orbital SE model. This model demonstrates, in agreement with the
SE model of Mostovoy and Khomskii \cite{Mos02} and with experiment
\cite{Rey01}, that the orbital SE $J_{\tau}$ is {\it stronger by one
order of magnitude\/} than any other (pure spin or spin-orbital)
interaction, because all magnetic dependence for the SE along
Ni--O--Ni bonds originates from the singlet-triplet splitting of the
oxygen $2p^4$ configuration, and is therefore smaller by at least a
factor $J_p/U_p\sim 0.1$, where $U_p$ ($J_p$)
is the interorbital Coulomb (Hund's exchange) interaction on oxygen.
These parameters, as well as the effective hopping $t$ and the charge
transfer energy $\Delta$, are of importance to establish quantitative
consequences of the SE model, in particular on such magnetic properties
as the effective exchange constants and the magnetic susceptibility.

The orbital SE has interesting consequences for the orbital
state and also for the magnetic interactions within the triangular
Ni planes. First of all, for an individual (NiO)$_2$ plaquette
(figure \ref{fig:pairorbs}), a pair of identical orbitals, either
both $d_{3z^2-r^2}+d_{x^2-y^2}$ or both $d_{3z^2-r^2}-d_{x^2-y^2}$, is
favoured. Therefore, in MF approximation a symmetry broken state
arises in the Ni planes, with FO order characterized by a single orbital
angle $\theta$ [as defined in equation (\ref{orbital})] along a
particular direction. This is in contrast with the behaviour obtained
for the cuprates \cite{Ole00} or the manganites \cite{Fei99}, where
orbitals with angles $\theta$ and $-\theta$ alternate on interlacing
sublattices. As a result,
one expects domains with either $|\theta\rangle$ or $|-\theta\rangle$
orbitals, separated by twin-boundaries. Even though much weaker than the
orbital SE, the spin-orbital SE plays a crucial role in the selection of
the ground state. It breaks the $U(1)$ symmetry generated in MF theory
by the orbital SE and restores the original threefold symmetry with
respect to the orbital angle $\theta$. By this mechanism FO order with
planar $d_{x^2-y^2}$ orbitals in the triangular Ni planes is stabilized
in MF theory. A characteristic feature of this state is the presence of
both FM and AF spin interactions.

After having a closer look at the magnetic properties of LiNiO$_2$,
we suggest that this compound represents an interesting case of
competition between the type of orbitals favoured locally by the
Jahn-Teller effect and those favoured by the SE in an $e_g$ system with
90$^{\circ}$ Ni-O-Ni bonds, and that this is likely to induce orbital
disorder.
Remarkably, this case is reminiscent of the situation encountered in the
$t_{2g}$ orbital model \cite{Kha01,vdB04}, and contradicts the
experience and common knowledge from the $e_g$ models in perovskites,
where the SE along 180$^{\circ}$ bonds favours alternating orbitals and
both mechanisms support each other \cite{Fei99,Wei04}.
Whether this competition might lead to dynamical disorder certainly
depends on the phonons, and we conclude that further study of elastic
coupling and phonons is urgently needed.

Summarizing, we presented and discussed the consequences of the 2D
{\it frustrated quantum spin-orbital model\/} derived for the
triangular Ni planes of two compounds with a very different behaviour:
LiNiO$_2$ and NaNiO$_2$. This model emphasizes the importance of
charge transfer (Goodenough) processes and presents a complete
description of the superexchange interactions on the 90$^{\circ}$
Ni-O-Ni bonds in triangular Ni planes, so we are confident that
it provides a solid starting point for future progress in the theory.
The frustration in the orbital sector is not of Ising
type, as suggested before \cite{Rey01}, but due to different orbitals
being favoured for different bond directions.
Providing a final answer concerning the origin of the different magnetic
properties of LiNiO$_2$ and NaNiO$_2$ requires an extension of
the present microscopic model by a microscopic description of both the
interlayer coupling and of the coupling between orbitals and the lattice.
We identified two possible reasons why the interlayer (spin and orbital)
coupling is so weak:
($i$) an intrinsic effect due to the small covalency of interplane
Ni--O--Li--O--Ni bonds, and
($ii$) an extrinsic effect due to Ni disorder in
       Li$_{1-x}$Ni$_{1+x}$O$_2$
--- both these effects are expected to play an important role and to be
responsible for the absence of $A$-type antiferromagnetic order in
LiNiO$_2$.

\ack
We thank D.I. Khomskii, G.A. Sawatzky and in particular
M. Dolores N\'u\~{n}ez-Regueiro and Jan Zaanen for stimulating
discussions. A.M.~Ole\'s would like to acknowledge support by
the Polish State Committee of Scientific Research (KBN) under
Project No.~1~P03B~068~26.

\appendix
\section{Derivation of the spin-orbital Hamiltonian}
\label{sec:appendix}

\begin{figure}
\vskip 2cm
\begin{center}
\includegraphics[width=7.0cm,angle=-90]{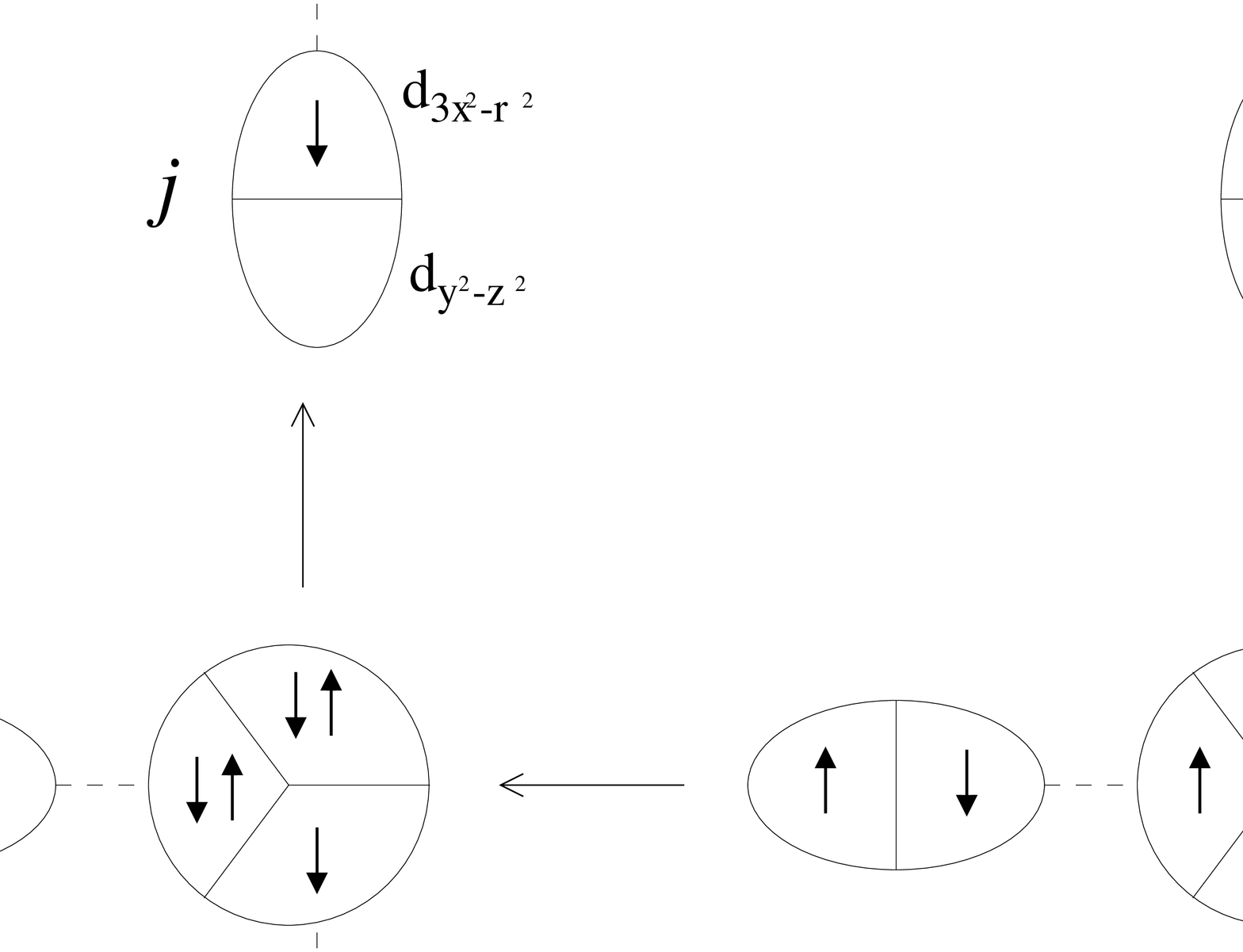}
\end{center}
\caption{
Example of superexchange processes (arrows in clockwise direction)
of $M$-type along the $XY$ direction. ---
After the first transition (top) either the triplet $T$ or the orbital
doublet $D$ is reached at what is now a Ni$^{2+}$ ion, while the second
transition (bottom right) gives either the orbital doublet $D$ or the
singlet $S$ at the other Ni$^{2+}$ ion. The corresponding double
excitation into the $2p^4$ configuration at oxygen is either to the
triplet $t$ or the singlet $s$.
--- Deexcitation occurs by the reverse transitions, either in the same
or in reversed order (shown in the diagram).
}
\label{fig:mtype}
\end{figure}

\begin{figure}
\vskip 1.7cm
\begin{center}
\includegraphics[width=7.0cm,angle=-90]{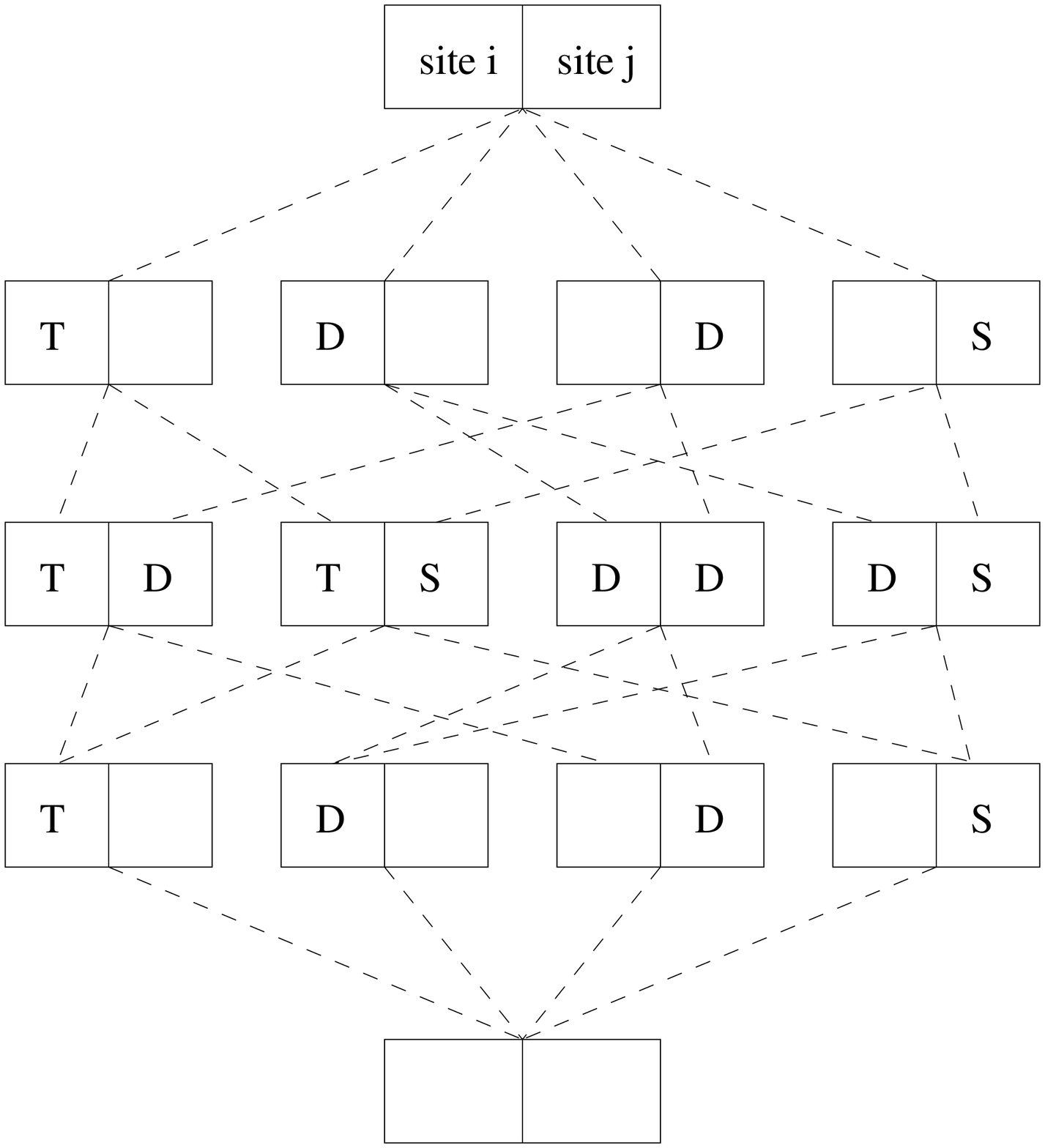}
\end{center}
\caption{
Diagrammatic representation of the superexchange processes shown in
figure \protect\ref{fig:mtype}, including inverted order of excitation
and deexcitation processes.
}
\label{fig:tree}
\end{figure}

Here we give further details on the derivation of the spin-orbital
Hamiltonian ${\mathcal H}_{\rm eff}$, in particular regarding the case
where $J_{\rm H}$ is finite.
First we take a closer look at the four-step charge transfer sequences.
As an example, let us consider (see figure \ref{fig:mtype}) a Ni--Ni
pair in the $XY$ direction in an M-type initial configuration with the
electron on Ni-site $i$ ($j$) in the non-hopping $d_{z^2-x^2}$ orbital
$|\bar{y}\rangle_i$ (hopping $d_{3x^2-r^2}$ orbital $|x\rangle_j$)
having spin up (down),
i.e. an initial state $|{\rm M}, \uparrow \downarrow \rangle$, and the
electrons being transferred from the oxygen having opposite spin.
In an obvious notation the corresponding contribution to
${\mathcal H}_{\rm eff}$ is
$
\lshad {\rm M},\uparrow \downarrow; \downarrow \uparrow \rshad \;
   {\mathcal P}_i^{\bar{y}}\, {\mathcal P}_j^{x} \:
    {\bi Q}_{ij}^{\uparrow \downarrow}
$.
In the middle intermediate state the ion states
$T$ and $D$, $t$ and $s$, and $D$ and $S$ occur with
equal amplitude at the Ni$^{2+}$, O$^{0}$ and Ni$^{2+}$ ions,
respectively.
Denoting the contribution from a doubly excited state by $[XvY]$, with
$X,Y \in \{T,D,S\}$ and $v \in \{t,s\}$, the contribution corresponding
to figure \ref{fig:mtype} is thus given by
$$
\lshad {\rm M}, \uparrow \downarrow; \downarrow \uparrow \rshad
  = \case{1}{2} \Big( [TtD] + [TtS] + [DtD] + [DtS]
                  + [TsD] + [TsS] + [DsD] + [DsS] \Big) ,
$$
where again a factor 4 has already been included. However, the four
sequences that lead to a particular middle state upon inverting the
order of the exciting and/or deexciting charge transfers, are now in
general inequivalent, as illustrated by the tree diagram of figure
\ref{fig:tree}, because the energies of the first and third intermediate
state may depend upon the sequence.
%
%
The contributions $[XvY]$ therefore need to be expanded as
\begin{equation}
\fl \hskip 1cm   [XvY] = \case{1}{4}
             \Big[(X|XvY|X)+(X|XvY|Y)+(Y|XvY|X)+(Y|XvY|Y) \Big],
\label{tree}
\end{equation}
where we denote in $(Z_1|XvY|Z_3)$ by $Z_1$ ($Z_3$) the state reached
after the first (third) charge transfer step.
%
%
%
%
The explicit expression for this quantity is [compare equation
(\ref{Good90})]
\begin{eqnarray}
\fl  (Z_1|XvY|Z_3) =
- \frac{t^4}{ \Delta_{Z_1} \Delta_{Z_3}
            ( \Delta_{X} + \Delta_{Y} + U_v ) }
+ \frac{t^4}{ \Delta_{Z_1} \Delta_{Z_3}
            ( \Delta_{X} + \Delta_{Y} )}  \nonumber \\
= \frac{t^4 \: U_v}{ \Delta_{Z_1} \Delta_{Z_3}
            ( \Delta_{X} + \Delta_{Y} )
            ( \Delta_{X} + \Delta_{Y} + U_v ) } ,
\label{ptgeneral}
\end{eqnarray}
where the second term in the first line is the renormalization
correction discussed in section \ref{sec:SEeg}.
Inserting (\ref{ptgeneral}) into equation (\ref{tree}) one now obtains
\begin{equation}
[XvY] =
 \frac{t^4}{4} \; \frac{ \Delta_{X} + \Delta_{Y} }
                       { \Delta_X^2 \: \Delta_Y^2 } \;
    \frac{U_v}{ \Delta_{X} + \Delta_{Y} + U_v } ,
\label{ptmiddle}
\end{equation}
so that for example
\begin{eqnarray}
[TsS] =
 \frac{ t^4 \: \Delta}{ 2 (\Delta^2 - 4 J_{\rm H}^2 )^2 }
 \;    \frac{U_p + J_p}{ 2 \Delta + U_p + J_p } ,     \\
\label{ptTsS}
[DtD] =
 \frac{ t^4 }{ 2  \Delta^3 }
  \;  \frac{U_p - J_p}{ 2 \Delta + U_p - J_p } .
\label{ptDtD}
\end{eqnarray}
%


The next step consists in listing all possible initial configurations
and for each one the allowed spin arrangements of the two hopping
electrons, and then working out, as above for
$\lshad {\rm M},\uparrow \downarrow; \downarrow \uparrow \rshad$,
with the help of diagrams like those shown in figure
\ref{fig:mtype}, which middle intermediate states contribute. The result
is shown in table \ref{table}, which gives the coefficients $C(..|..)$
in
\begin{equation}
\lshad {\rm L},\sigma \sigma'; \bar{\sigma} \bar{\sigma}'\rshad
  = \sum_{X,Y,v} C({\rm L},\sigma \sigma'; \bar{\sigma} \bar{\sigma}'|
                   XvY) \, [XvY] .
\label{matrix}
\end{equation}
The SE coupling constants $\{ K_{\rm L}^{\rm T,S}\}$ for spin triplet 
and spin singlet, appearing in equation (\ref{HamK}), are now obtained 
as follows. In each case (L = O, M or N) consider first the $m=1$ 
component of the triplet, i.e. the row(s) in table \ref{table}
in which there is an up-spin on both Ni$^{3+}$ ions.
Adding them gives the triplet contribution $K_{\rm L}^{\rm T}$, while
similarly adding the row(s) for which the two Ni$^{3+}$ spins are
in up-down configuration gives $K_{\rm L}^{\uparrow \downarrow}$,
\begin{eqnarray}
 K_{\rm L}^{\rm T} = \sum_{\bar{\sigma}, \bar{\sigma}'}
   \lshad {\rm L},\uparrow \uparrow; \bar{\sigma} \bar{\sigma}'\rshad
 = \sum_{\bar{\sigma}, \bar{\sigma}'} \;  \sum_{X,Y,v}
   C({\rm L}, \uparrow \uparrow; \bar{\sigma} \bar{\sigma}'|
                                             XvY) \, [XvY] , \\
 K_{\rm L}^{\uparrow \downarrow} = \sum_{\bar{\sigma}, \bar{\sigma}'}
   \lshad {\rm L},\uparrow \downarrow; \bar{\sigma} \bar{\sigma}'\rshad
 = \sum_{\bar{\sigma}, \bar{\sigma}'} \;  \sum_{X,Y,v}
   C({\rm L}, \uparrow \downarrow; \bar{\sigma} \bar{\sigma}'|
                                             XvY) \, [XvY] .
\label{K's}
\end{eqnarray}
As $K_{\rm L}^{\uparrow \downarrow}$ corresponds to the $m=0$
components and so is shared between singlet and triplet, one now
obtains the singlet contribution from
$K_{\rm L}^{\rm S} = 2 K_{\rm L}^{\uparrow \downarrow} 
                     - K_{\rm L}^{\rm T}$.
%
%
%
\begin{table}
\caption{\label{table} Coefficients
$C({\rm L},\sigma \sigma'; \bar{\sigma} \bar{\sigma}'|XvY)$
specifying the contributions to the superexchange Hamiltonian according
to equation (\protect\ref{matrix}). Here ${\rm L},\sigma \sigma'$
denotes the initial configuration and $\bar{\sigma} \bar{\sigma}'$
the spins of the transferred electrons, with
${\rm L} \in \{ {\rm O}, {\rm M}, {\rm N} \}$ and
$\sigma, \sigma', \bar{\sigma}, \bar{\sigma}'
\in \{ \uparrow, \downarrow \}$,
and $XvY$ denotes the middle intermediate state in the four-step charge
transfer process, with
$X,Y \in \{T,D,S\}$ and $v \in \{t,s\}$.
}
\begin{indented}
\item[]\begin{tabular}{@{}ccccccccccccc}
\br
    & $ \!\! TtT \! $ & $ \!\! TsT \! $
    & $ \!\! TtD \! $ & $ \!\! TsD \! $
    & $ \!\! TtS \! $ & $ \!\! TsS \! $
    & $ \!\! DtD \! $ & $ \!\! DsD \! $
    & $ \!\! DtS \! $ & $ \!\! DsS \! $
    & $ \!\! StS \! $ & $ \!\! SsS \! $ \\
\mr
O,$\uparrow \uparrow;\downarrow \downarrow$
    &  0  &  0  &  0  &  0  &  0  &  0  &  1  &  0  &  2  &  0
                                                    &  1  &  0   \\
O,$\uparrow \downarrow;\downarrow \uparrow$
    &  0  &  0  &  0  &  0  &  0  &  0  & $\case{1}{2}$ & $\case{1}{2} $
          &  1  &  1  & $\case{1}{2}$ & $\case{1}{2}$  \\
\mr
M,$\uparrow \uparrow;\uparrow \downarrow$
    &  0  &  0  &  1  &  1  &  1  &  1  &  0  &  0  &  0  &  0
                                                    &  0  &  0   \\
M,$\uparrow \uparrow;\downarrow \downarrow$
    &  0  &  0  &  1  &  0  &  1  &  0  &  1  &  0  &  1  &  0
                                                    &  0  &  0   \\
M,$\uparrow \downarrow;\uparrow \uparrow$
    &  0  &  0  &  2  &  0  &  2  &  0  &  0  &  0  &  0  &  0
                                                    & 0  &  0   \\
M,$\uparrow \downarrow;\downarrow \uparrow$
    &  0  &  0  & $\case{1}{2}$ & $\case{1}{2}$ & $\case{1}{2}$ &
  $\case{1}{2}$ & $\case{1}{2}$ & $\case{1}{2}$ & $\case{1}{2}$ &
  $\case{1}{2}$ &  0  &  0   \\
\mr
N,$\uparrow \uparrow;\uparrow \uparrow$
    &  4  &  0  &  0  &  0  &  0  &  0  &  0  &  0  &  0  &  0
                                                    &  0  &  0   \\
N,$\uparrow \uparrow;\uparrow \downarrow$
    &  1  &  1  &  1  &  1  &  0  &  0  &  0  &  0  &  0  &  0
                                                    &  0  &  0   \\
N,$\uparrow \uparrow;\downarrow \uparrow$
    &  1  &  1  &  1  &  1  &  0  &  0  &  0  &  0  &  0  &  0
                                                    &  0  &  0   \\
N,$\uparrow \uparrow;\downarrow \downarrow$
    &  1  &  0  &  2  &  0  &  0  &  0  &  1  &  0  &  0  &  0
                                                    &  0  &  0   \\
N,$\uparrow \downarrow;\uparrow \uparrow$
    &  2  &  0  &  2  &  0  &  0  &  0  &  0  &  0  &  0  &  0
                                                    &  0  &  0   \\
N,$\uparrow \downarrow;\uparrow \downarrow$
    &  2  &  2  &  0  &  0  &  0  &  0  &  0  &  0  &  0  &  0
                                                    &  0  &  0   \\
N,$\uparrow \downarrow;\downarrow \uparrow$
    & $\case{1}{2}$ & $\case{1}{2}$ &  1  &  1  &  0  &  0  &
      $\case{1}{2}$ & $\case{1}{2}$ &  0  &  0  &  0  &  0   \\
N,$\uparrow \downarrow;\downarrow \downarrow$
    &  2  &  0  &  2  &  0  &  0  &  0  &  0  &  0  &  0  &  0
                                                    &  0  &  0   \\
\br
\end{tabular}
\end{indented}
\end{table}
%
%
The result is
\begin{eqnarray}
\label{KTsO}
K_{\rm O}^{\rm T} = [DtD] + 2 [DtS] + [StS] ,  \\
\label{KSsO}
K_{\rm O}^{\rm S} = [DsD] + 2 [DsS] + [SsS] ,  \\
\label{KTsM}
K_{\rm M}^{\rm T} = 2 [TtD] + [TsD] + 2 [TtS] + [TsS]
+ [DtD] + [DtS] ,
        \\
\label{KSsM}
K_{\rm M}^{\rm S} = 3 [TtD] \hskip 1.60cm + 3 [TtS] \hskip 1.50cm
+ [DsD] + [DsS] ,
        \\
\label{KTsN}
K_{\rm N}^{\rm T} = 7 [TtT] + 2 [TsT] + 4 [TtD] + 2 [TsD]
+ [DtD] ,
        \\
\label{KSsN}
K_{\rm N}^{\rm S} = 6 [TtT] + 3 [TsT] + 6 [TtD]  \hskip 1.80cm
+ [DsD] .
\end{eqnarray}
The above procedure (considering first the $m=1$ component of the
triplet) avoids the necessity of keeping track of the phases.
Alternatively one may work out explicitly the states resulting
after two perturbation steps from both
$|{\rm L}, \uparrow \downarrow \rangle$ and
$|{\rm L}, \downarrow \uparrow \rangle$,
and from those, by taking their sum and difference, the states resulting
from the $m=0$ components of the triplet and the singlet, and finally
project the latter on all possible middle states. It is
straightforward to verify that this gives the same results
and equations (\ref{KTsO})--(\ref{KSsN}) are reproduced.
One should note that, while the parallel-spin initial state
$|{\rm L}, \uparrow \uparrow \rangle$ corresponds necessarily to the
triplet and is therefore associated with FM coupling, it is
not correct to associate the antiparallel-spin initial state
$|{\rm L}, \uparrow \downarrow \rangle$ with AF
coupling, since it projects on both the triplet and the singlet.
%

The SE constants $J_{\rm L}^{0,S}$ occurring in equation 
(\ref{HamJ}) then follow from equations (\ref{JinK}) and 
(\ref{KTsO})--(\ref{KSsN}), with the result
\begin{eqnarray}
\label{J0sO}
\fl  J_{\rm O}^{0} = \hskip 5.70cm
                       (DD)_{+}    + 2 (DS)_{+} +   (SS)_{+}  ,
\\
\label{J0sM}
\fl  J_{\rm M}^{0} = \hskip 1.85cm   3 (TD)_{+} + 3 (TS)_{+}
                   +   (DD)_{+} + \hskip 0.22cm     (DS)_{+} ,
\\
\label{J0sN}
\fl  J_{\rm N}^{0} = 9 (TT)_{+}    + 6 (TD)_{+}   \hskip 1.90cm
                   +   (DD)_{+}   ,
\end{eqnarray}
\begin{eqnarray}
\label{JSsO}
\fl  J_{\rm O}^{S} = \hskip 5.30cm
                       (DD)_{-}    + 2 (DS)_{-} + (SS)_{-}  ,
\\
\label{JSsM}
\fl  J_{\rm M}^{S} = \hskip 1.22cm - \hskip 0.30 cm   (TD)_{-}
                   - (TS)_{-}
                   + (DD)_{-} + \hskip 0.22cm  (DS)_{-} ,
\\
\label{JSsN}
\fl  J_{\rm N}^{S} = (TT)_{-}    - 2 (TD)_{-}
                     \hskip 1.70cm + (DD)_{-} \,  ,
\end{eqnarray}
where we have introduced the following abbreviations for the orbital
exchange elements occurring in (\ref{J0sO})--(\ref{J0sN}) and the
spin-orbital exchange elements occurring in (\ref{JSsO})--(\ref{JSsN}),
\begin{eqnarray}
\label{abbrev:a}
(XY)_{+} &=& \case{3}{4} [XtY] + \case{1}{4} [XsY],   \\
\label{abbrev:b}
(XY)_{-} &=&  [XtY] - [XsY].
\end{eqnarray}
One readily verifies that equations (\ref{J0sO})--(\ref{JSsN}) reduce to
equations (\ref{J_0})--(\ref{J_TS}) upon setting $J_{\rm H}=0$,
whereupon all $(XY)_{+}$ become identical as do all $(XY)_{-}$.


From equations (\ref{abbrev:a})--(\ref{abbrev:b}) or from the explicit
expressions
\begin{eqnarray}
(XY)_{+} &=&
    \frac{t^4}{4} \; \frac{ \Delta_{X} + \Delta_{Y} }
                          { \Delta_X^2 \: \Delta_Y^2 } \;
    \frac{ ( U_p - \case{1}{2} J_p ) ( \Delta_{X} + \Delta_{Y} )
                          + U_p^2 - J_p^2 }
         { ( \Delta_{X} + \Delta_{Y} + U_p )^2 - J_p^2 } ,  \\
(XY)_{-} &=&
 -  \frac{t^4}{2} \; \frac{ ( \Delta_{X} + \Delta_{Y} )^2 }
                          { \Delta_X^2 \: \Delta_Y^2 } \;
    \frac{ J_p }
         { ( \Delta_{X} + \Delta_{Y} + U_p )^2 - J_p^2 } ,
\label{SEels}
\end{eqnarray}
it is obvious that the orbital exchange elements $\{ (XY)_{+} \}$
are much larger than the spin-orbital exchange elements
$\{ (XY)_{-} \}$, the ratio being
\begin{equation}
\frac{ |(XY)_{-}| }{ (XY)_{+} } =
    \frac{ 2 J_p \: ( \Delta_X + \Delta_Y ) }
         { ( U_p - \case{1}{2} J_p ) ( \Delta_X + \Delta_Y )
           + U_p^2 - J_p^2 }
  \simeq \frac{ 4 \Delta \:  J_p }
              { ( 2 \Delta + U_p ) \: U_p } ,
\label{(XY)ratio}
\end{equation}
where the second expression applies for
$J_p, J_{\rm H} \ll \Delta , U_p$. It then follows from equations
(\ref{J0sO})--(\ref{J0sN}) and (\ref{JSsO})--(\ref{JSsN}) that the same
holds for the orbital SE constant $J_{\tau}$ [or $\bar{J}'_T$, see
equation (\ref{HamT})] when compared with any of the spin or
spin-orbital SE constants $J_{\sigma}$, $J_{\nu}$ or $J_{\mu}$
[or $\bar{J}_{TS}$, $\bar{J}'_{TS}$, or $\bar{J}''_{TS}$, see
equation (\ref{HamTS})]
--- the same ratio (\ref{(XY)ratio}) applies to
$\bar{J}_{TS} / \bar{J}'_{T}$, in accordance with the analysis made
above for $J_{\rm H}=0$ [see equations (\ref{J_T})--(\ref{J_TS})].

As for the relative size of the spin and spin-orbital constants with
respect to one another, we observe that while $J_{\rm O}^S$ is the sum
of four spin-orbital exchange elements [equation (\ref{JSsO})],
$J_{\rm M}^S$ is the sum of two differences of such elements, while
$J_{\rm N}^S$ is the difference of two differences of such elements,
\begin{equation}
J_{\rm M}^{S} =  D_{S} + D_{D} , \hskip 1cm
J_{\rm N}^{S} =  D_{D} - D_{T} ,
\label{sumdiff}
\end{equation}
with
\begin{equation}
\label{sumdiffX}
  D_{X} = (DX)_{-} - (TX)_{-}
  \simeq \Big( \frac{1}{\Delta_T} - \frac{1}{\Delta_D} \Big)
     ( \frac{1}{\Delta_T} + \frac{1}{\Delta_D}
                          + \frac{2}{\Delta_X} \Big)  C_{TDX} ,
\end{equation}
where $C_{TDX}$ depends only weakly upon $J_{\rm H}$ and is given
in good approximation by
\begin{equation}
\label{C-TDX}
C_{TDX}
 \simeq  \frac{ t^4 }{ 2 }
 \;    \frac{J_p}{ ( 2 \Delta + U_p )^2 - J_p^2 }    .
\end{equation}
Obviously all $D_X$ are positive since $ \Delta_T < \Delta_D $, and
further $ D_T > D_D > D_S $ since $ \Delta_T < \Delta_D < \Delta_S $.
It follows that $ J_{\rm M}^S > 0 $ and $ J_{\rm N}^S < 0 $, and
moreover, as clearly $ D_S + 2 D_D $ is considerably larger than $D_T$,
one concludes that
\begin{equation}
 |J_{\rm O}^S|  \gg J_{\rm M}^S  \gg |J_{\rm N}^S|  .
\label{Jrelationapp}
\end{equation}

Finally we arrive at the SE constants associated with the more physical
interactions as occurring in ${\mathcal H}_{\rm eff,o}$ [equation
(\ref{HamT})] and ${\mathcal H}_{\rm eff,s}$ [equation
(\ref{HamTS})].
The orbital part is described by
\begin{equation}
J_{\tau}
= \frac{1}{2} (J_{\rm O}^0 - 2 J_{\rm M}^0 + J_{\rm N}^0 )
= \frac{1}{2} [ 9 (TT)_{+} - 6 (TS)_{+} + (SS)_{+} ] ,
\label{Jgreekorb}
\end{equation}
while the pure spin SE constant $J_{\sigma}$ and the two spin-orbital SE
constants $J_{\nu}$ and $J_{\mu}$ are given by
\begin{eqnarray}
\fl  J_{\sigma}
= - \frac{1}{2} ( J_{\rm O}^S + 2 J_{\rm M}^S + J_{\rm N}^S )
\nonumber  \\
\lo = - \frac{1}{2} [   (TT)_{-} - 4 (TD)_{-} - 2 (TS)_{-}
                    + 4 (DD)_{-} + 4 (DS)_{-} + (SS)_{-} ] ,
                        \\
\fl J_{\mu}
= - \frac{1}{2} ( J_{\rm O}^S - 2 J_{\rm M}^S + J_{\rm N}^S )
\nonumber  \\
\lo = - \frac{1}{2} [(TT)_{-} + 2 (TS)_{-} +  (SS)_{-} ]
\nonumber \\
\lo = J_{\sigma} + 2 D_S + 2 D_D ,
                \\
\fl J_{\nu}
= - \frac{1}{2} (J_{\rm O}^S-J_{\rm N}^S)
\nonumber  \\
\lo = - \frac{1}{2} [- (TT)_{-} + 2 (TD)_{-} + 2 (DS)_{-}
+ (SS)_{-} ]
\nonumber \\
\lo = J_{\sigma} + D_S + 2 D_D -D_T .
\label{Jgreekspin}
\end{eqnarray}
From these expressions and the inequality (\ref{Jrelationapp})
above, the inequality (\ref{Jinequality}) follows immediately.


\end{document}